\documentclass[twocolumn]{jpsj2}

\title{%
Fourth-Order Perturbation Expansion for Hubbard Model
on a Two-Dimensional Square Lattice 
}

\author{%
Hiroaki \textsc{Ikeda}$^1$
\thanks{E-mail:hiroaki@scphys.kyoto-u.ac.jp},
Shogo \textsc{Shinkai},
and Kosaku \textsc{Yamada}$^2$
}

\inst{
$^1$Department of Physics, Kyoto University, Kyoto 606-8502 \\
$^2$Faculty of Science and Engineering, Ritsumeikan University, 
Kusatsu 525-8577
}

\recdate{Received March 28, 2007; accepted March 14, 2008; published May 26, 2008}

\abst{%
We investigate the Hubbard model on a two-dimensional square lattice by the perturbation expansion to the fourth order in the on-site Coulomb repulsion $U$.
Numerically calculating all diagrams up to the fourth order in self-energy, we examine the convergence of perturbation series in the lattice system.
We indicate that the coefficient of each order term rapidly decreases as in the impurity Anderson model for $T \gtrsim 0.1t$ in the half-filled case, but it holds in the doped case even at lower temperatures.
Thus, we can expect that the convergence of perturbation expansion in $U$ is very good in a wide parameter region also in the lattice system, except for $T\lesssim 0.1t$ in the half-filled case.
We next calculate the density of states in the fourth-order perturbation.
In the half-filled case, the shape in a moderate correlation regime is quite different from the three peak structure in the second-order perturbation.
Remarkable upper and lower Hubbard bands locate at $\omega\simeq\pm U/2$, and a pseudogap appears at the Fermi level $\omega=0$.
This is considered as the precursor of the Mott-Hubbard antiferromagnetic structure.
In the doped case, quasiparticles with very heavy mass are formed at the Fermi level.
Thus, we conclude that the fourth-order perturbation theory overall well explain the asymptotic behaviors in a strong correlation regime.
}

\kword{%
Fermi liquid theory, perturbation expansion, Fourth-order perturbation,
Hubbard model, Mott-Hubbard band, pseudogap, Fermi arc
}

\begin{document}
\maketitle

\section{Introduction}
Over the past several decades, research on high-$T_\mathrm{c}$ cuprates has been a major theme in condensed matter physics. 
Unconventional superconductivity with line nodes, not an isotropic gap in the BCS theory, appears with carrier doping in the Mott insulator. 
Inevitably, a strong correlation has been considered very important. 
The ground state in the Mott insulator is the antiferromagnetic (AF) state, and the interplay between magnetism and unconventional superconductivity has been the key issue. 
In order to understand the phase diagram and physical properties, a large number of theoretical and experimental research studies have been performed.~\cite{rf:Rev1,rf:Rev2}
Today, we understand that superconductivity originates from the AF spin fluctuation. 
On the same footing, most anomalous features in cuprates can be consistently explained.~\cite{rf:Kontani,rf:Kontani2}

Quantitatively, fluctuation-exchange approximation (FLEX) has been used as an efficient numerical method.
This method successfully describes the behavior of the Fermi liquid (FL) state in the vicinity of the AF critical point on the basis of the microscopic Hamiltonian.
However, it has a few disadvantages.
The specific mode of the spin fluctuation is overestimated owing to the enhancement by the partial summation of bubble and ladder diagrams.
The Hubbard peak structure in the density of states (DOS) is smeared and unclear.
The Mott transition cannot be described.
These failures in the normal state are concerned with the fact that the Mott-Hubbard character arises from a local correlation.
The essence has been clarified by dynamical mean-field theory (DMFT).~\cite{rf:Zhang,rf:Georges}
Nowadays, great efforts have been made to take short-range correlations into account,~\cite{rf:Maier,rf:Maier2,rf:Kyung,rf:Sadovskii,rf:Kusunose,rf:Toschi,rf:Imada} such as the dynamical cluster approximation (DCA).
As for superconductivity, on the other hand, the FLEX well explains the d-wave spin-singlet superconductivity near the AF phase, and the p-wave spin-triplet channel can be dominant near the ferromagnetic phase.
These are reasonable results.
However, the FLEX cannot simply explain the p-wave spin-triplet superconductivity in Sr$_2$RuO$_4$.
In order to obtain the triplet state in a single-band Hubbard model for the $\gamma$ band, which is considered to trigger the superconductivity in Sr$_2$RuO$_4$, an additional off-site interaction is necessary at least.~\cite{rf:Arita}
This may be related to the fact that the spin fluctuation in the $\gamma$ band shows a featureless broad hump at around the $\mathit{\Gamma}ma$ point, although the FLEX is justified in systems with remarkable spin fluctuation.
There is a possibility that the FLEX ignores crucial terms in the p-wave state in Sr$_2$RuO$_4$.
In fact, the third-order perturbation theory~\cite{rf:Nomura,rf:Nomura2} and the one-loop renormalization group method~\cite{rf:Honerkamp,rf:Honerkamp2} indicate a p-wave state within a conventional single-band Hubbard model for the $\gamma$ band.
In these cases, the important process in the pairing interaction mainly originates from the third-order term.
This is the case for the spin-triplet channel in the two-dimensional electron gas system.~\cite{rf:Chubukov,rf:Feldman}
In the FLEX, the process, which is classified into a vertex correction term, is not included.
Thus, the FLEX may not work except in the vicinity of the critical point such as the AF (or ferromagnetic) phase, although we need further investigation to clarify it.

On the other hand, the third-order perturbation theory seems to be efficient in the study of superconductivity.
It is comprehensively applicable in both spin singlet and triplet states.
Input parameters are only dispersion relation and carrier number.
However, this method is an approach based on a weak correlation.
The applicability in strongly correlated systems is obscure.
We need to evaluate the convergence of the perturbation expansion.
In addition, we can expect that a very large mass enhancement factor is obtained in higher-order perturbation from previous studies.~\cite{rf:Shinkai,rf:Shinkai2}
Thus, higher-order perturbation in the on-site Coulomb repulsion $U$ is important not only as fundamental knowledge in strongly correlated systems, but also as a quantitative method.
Up to now, from numerical calculations restricted to ${\mib k}$ points on the Fermi surface (FS), Nomura and Yamada have discussed the convergence of the pairing interaction in the fourth-order perturbation,~\cite{rf:Nomura3} and Shinkai et. al have evaluated the mass renormalization factor.~\cite{rf:Shinkai,rf:Shinkai2}
However, owing to the restriction, physical properties in the fourth-order perturbation have not been clarified yet.
Thus, in this study, we carry out the perturbation expansion to the fourth order in the entire first Brillouin zone.
Its calculation is very instructive, and clarifies physical properties that are indefinite in previous research studies.~\cite{rf:Shinkai,rf:Shinkai2,rf:Nomura3,rf:Zlatic}

In the present study, one of theoretical backgrounds is the result of the perturbation approach of Yamada and Yosida for the impurity Anderson model.~\cite{rf:Yamada,rf:Yamada2,rf:Yosida}
They carried out the perturbation expansion to the fourth order in $U$ in the context of the Kondo problem.
Today, we have the exact solution, and the applicability of the perturbation theory has been confirmed.~\cite{rf:Zlatic2}
The radius of convergence is infinite, and physical quantities are analytic in $U$.
For instance, the spin susceptibility and the specific heat coefficient  are enhanced like the exponential function as a function of $U$.
Thus, the coefficient of each $n$th-order term in these physical quantities rapidly decreases almost in proportion to $\sim 1/n!$.
Because of this remarkable property, even if we truncate the perturbation expansion at a finite order, physical quantities rapidly approach the exact values with increasing cutoff order such as $2$, $4$, $6$, $\cdots$.
For instance, the exact Wilson ratio is $\sim 1.962$ at $u=U/\pi\mathit{\Delta}ta=2$.~\cite{rf:Zlatic2}
This can be regarded as a sufficiently strong correlation regime, since the exact value is $2$ at a strong correlation limit.
In this case, the approximate values are $\sim 1.639$ for the second order, $\sim 1.889$ for the fourth order and $\sim 1.952$ for the sixth order.
This indicates that the fourth-order perturbation expansion has sufficient accuracy in a moderate correlation regime.
Such good convergence is one of characteristics in the FL state.

Since we have no exact solution in two- or higher-dimensional lattice systems, we cannot guarantee the convergence of the perturbation expansion.
In fact, the ground states in many lattice systems are not the FL state, but the magnetic or superconducting state.
We will need much higher-order perturbation terms to restore the critical fluctuation near these critical points.
It may be rather better to perform partial summations, such as the FLEX.
However, above these transition temperatures, the system can be considered to be in the FL state, because of the principle of adiabatic continuation stressed by Anderson.~\cite{rf:Anderson}
As long as no phase transition occurs, the system connects adiabatically with the noninteracting system, and stays in the FL state.
In this case, the perturbation expansion is still applicable.
The physical quantities of the system will asymptotically approach the exact behavior with increasing cutoff order in the perturbation expansion.
In this paper, we perform the perturbation expansion for the normal self-energy.
Since the first-order term provides only a constant shift, the second-order term is the first one that includes correlation effects due to the on-site Coulomb repulsion $U$.
The next significant term comes from the fourth-order term.
This is because the third-order term is relatively small owing to the fact that it vanishes in systems with particle-hole symmetry.
The fourth-order term can include correlation effects that have not been grasped in the third-order perturbation theory so far.
This can qualitatively change the asymptotic behavior in the strong correlation regime.
In addition, we can estimate the validity of the perturbation expansion to the third order by comparing each order term.
If the perturbation expansion has good convergence of $1/n!$ as in the impurity case, the fourth-order perturbation theory will be valid in a wider parameter region than the third-order perturbation.
Thus, the fourth-order perturbation theory in lattice systems can be considered as one of several efficient methods of studying strongly correlated systems.
The investigation of the fourth-order perturbation is very fruitful not only as fundamental knowledge in strongly correlated systems, but also as a quantitative method.

In this paper, we calculate the fourth-order term for the normal self-energy in lattice systems, and examine the convergence of the perturbation expansion.
In addition, we investigate single-particle quantities, such as the DOS, in the fourth-order perturbation theory.
In particular, we clarify the asymptotic behavior in the strong correlation regime.
It is meaningful that we compare these results with those given by many other theoretical studies.
Furthermore, it will be useful to introduce numerical algorithms to evaluate vertex correction terms, which do not have the convolution form.

This paper is organized as follows.
In \S 2, we briefly formalize the perturbation expansion to the fourth order for the self-energy.
In \S 3, we describe some techniques in numerical calculations, particularly how to estimate vertex correction terms, which are usually neglected because of not being in the convolution form.
We present numerical results in both the half-filled and doped cases in \S 4.
We discuss the convergence of the perturbation expansion in the former half, and then show properties of single-particle quantities in the latter half.
Finally, in \S 5, we summarize our study and give an outline of future works.

\section{Formalism}
In this paper, we study the Hubbard Hamiltonian on an $N\times N$ square 
lattice:
\begin{equation}
H=\sum_{{\mib k}\sigma} \bigl(\epsilon_{\mib k}-\mu\bigr) 
c_{{\mib k}\sigma}^\dagger c_{{\mib k}\sigma}^{\vphantom{\dagger}}
+U\sum_i n_{i\uparrow}n_{i\downarrow}.
\end{equation}
$U$ is the on-site Coulomb repulsion, and $\mu$ is the chemical potential.
The dispersion relation $\epsilon_{\mib k}$ is given by 
\begin{equation}
\epsilon_{\mib k}=-2t\bigl(\cos(k_x)+\cos(k_y)\bigr),
\label{eq:Disp}
\end{equation}
where the wave vector ${\mib k}=(k_x,k_y)$ is measured in units of the inverse of lattice constant.
The single-particle Green's function is expressed with the self-energy $\mathit{\Sigma}(k)$ as
\begin{equation}
\mathcal{G}(k)^{-1}=\mathcal{G}_0(k)^{-1}-\mathit{\Sigma}(k)=\mathrm{i}\omega_n-\xi_{\mib k}-\mathit{\Sigma}(k),
\label{eq:Dyson}
\end{equation}
where $\xi_{\mib k}=\epsilon_{\mib k}-\mu$ and $k=({\mib k},\mathrm{i}\omega_n)$.
$\omega_n=(2n+1)\pi T$ is the fermion Matsubara frequency at a temperature $T$.
Here, we set the Boltzmann constant $k_mathrm{B}=1$.
Hereafter, we measure energy in units of $t$, that is, we set $t=1$.
For a given chemical potential $\mu$, the electron density $n$ is determined by
\begin{equation}
n=\sum_\sigma\left<n_{i\sigma}\right>=\sum_{k\sigma}\mathcal{G}_0(k)
=\frac{1}{N^2}\sum_{{\mib k}\sigma} f(\xi_{\mib k}),
\end{equation}
in the noninteracting system, and
\begin{equation}
n=\sum_{k\sigma}\mathcal{G}(k)=
\sum_{k\sigma}\Bigl(\mathcal{G}(k)-\mathcal{G}_0(k)\Bigr)+\frac{1}{N^2}\sum_{{\mib k}\sigma} f(\xi_k),
\label{eq:nmu}
\end{equation}
in the interacting system.
In these final expressions,
\begin{equation}
f(\epsilon)=\frac{1}{\mathrm{e}^{\beta\epsilon}+1}
=\frac{1}{2}\biggl(1-\tanh\Bigl(\frac{\beta\epsilon}{2}\Bigr)\biggr),
\label{eq:FD}
\end{equation}
is the Fermi-Dirac distribution function with $\beta=1/T$, and the sum over $k$ denotes
\begin{equation*}
\sum_k=\frac{T}{N^2}\sum_{\mib k}\sum_{\mathrm{i}\omega_n}.
\end{equation*}
The final expressions in eqs.~(\ref{eq:nmu}) and (\ref{eq:FD}) are convenient in numerical calculations.

\begin{figure}[t]
\begin{center}
\vspace{20 pt}
\includegraphics[width=6.5cm]{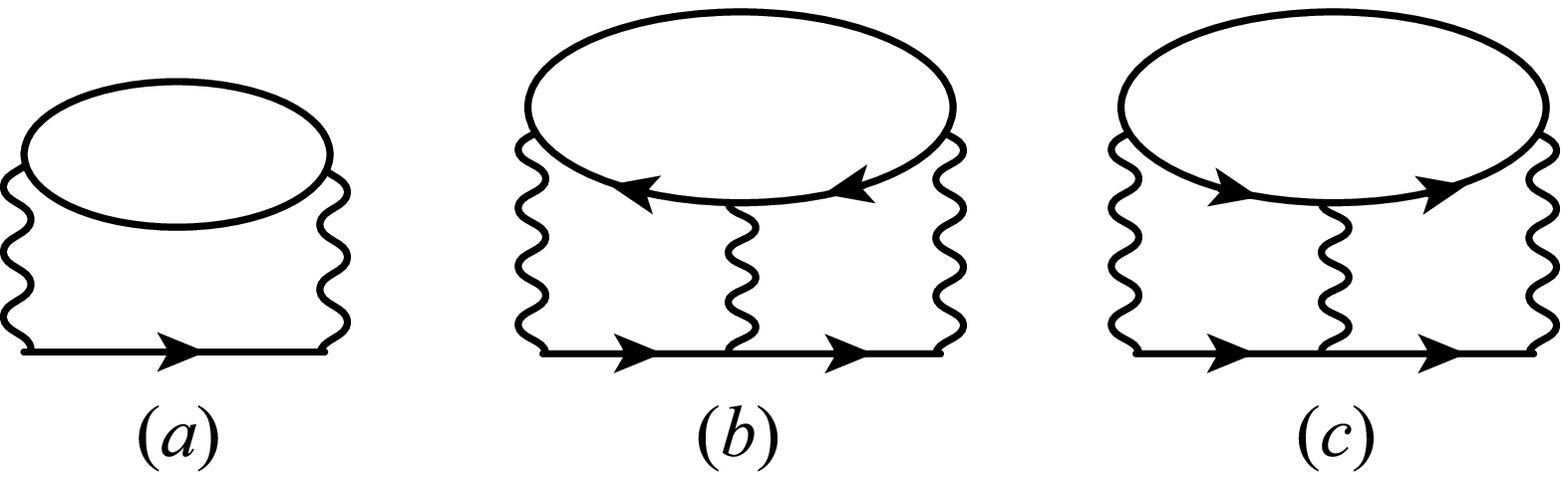}
\caption{Diagrams of the second-order term $\mathit{\Sigma}^{(2)}(k)$ and the third-order term $\mathit{\Sigma}^{(3)}(k)$ in self-energy.}
\label{fig:sig3}

\vspace{20 pt}
\includegraphics[width=7cm]{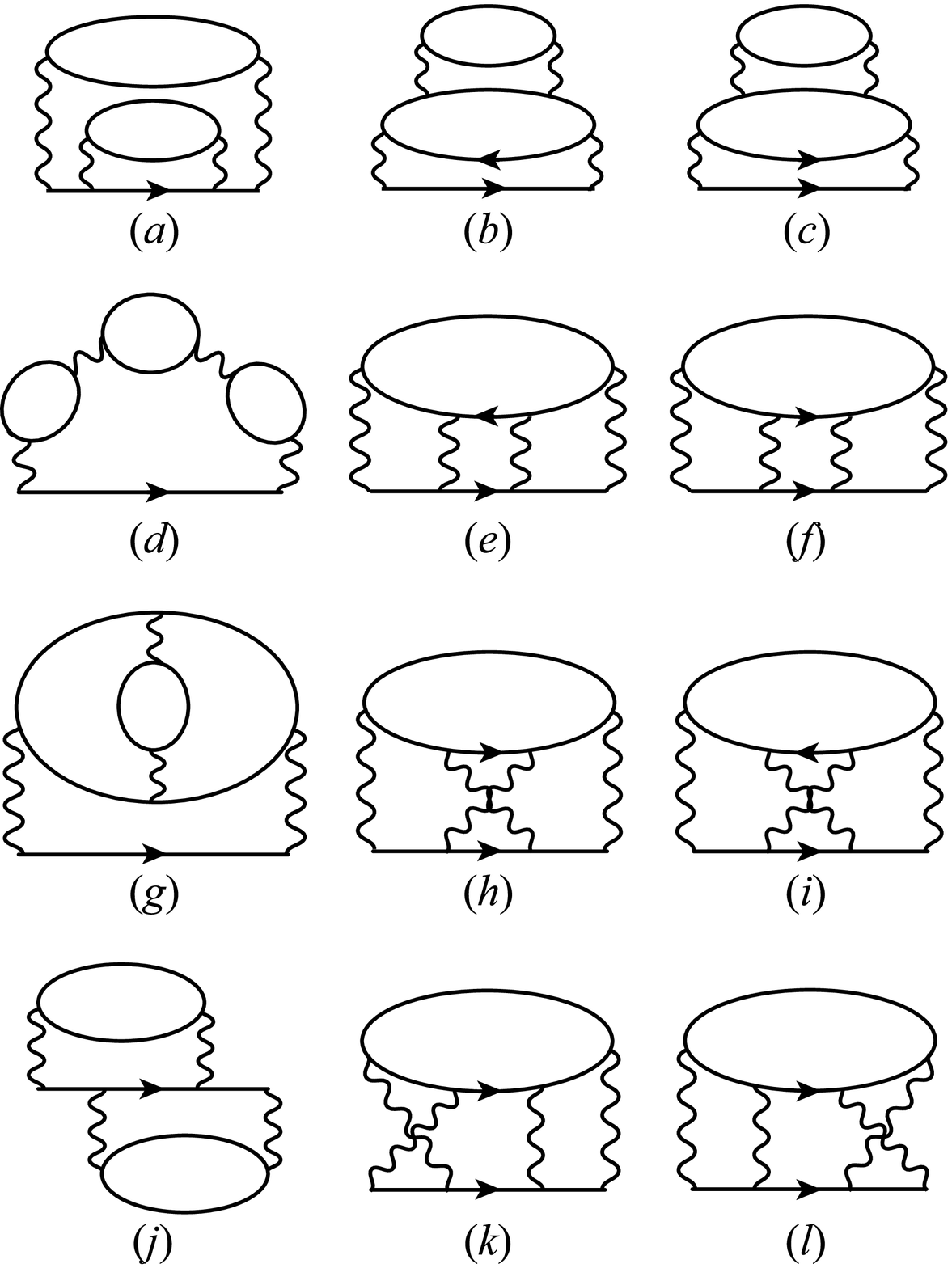}
\end{center}
\caption{Diagrams of the fourth-order term $\mathit{\Sigma}^{(4)}(k)$ in self-energy.
$(a)-(c)$ correspond to $\mathit{\Sigma}^{(4)}_\mathrm{self}(k)$, which is the reducible self-energy including Green's function with the second-order self-energy;
$(d)-(f)$ $\mathit{\Sigma}^{(4)}_\mathrm{RPA}(k)$, which is the self-energy mediated by one boson fluctuation;
$(g)-(i)$ $\mathit{\Sigma}^{(4)}_\mathrm{vtx1}(k)$, which denotes the vertex correction for one boson fluctuation itself;
$(j)-(l)$ $\mathit{\Sigma}^{(4)}_\mathrm{vtx2}(k)$, which denotes the vertex correction with two boson fluctuation crossing.}
\label{fig:sig4}
\end{figure}
In this study, we carry out the perturbation expansion to the fourth order in $U$ for the self-energy $\mathit{\Sigma}(k)$.
The self-energy $\mathit{\Sigma}(k)$ in $u=U/t$ is expanded as 
\begin{equation}
\mathit{\Sigma}(k)=\mathit{\Sigma}^{(2)}(k)u^2+\mathit{\Sigma}^{(3)}(k)u^3+\mathit{\Sigma}^{(4)}(k)u^4,
\end{equation}
where we neglect the first-order term, that is, the Hartree term.
This term provides only a constant shift, and can then be included by the chemical potential shift.
All diagrams that appear in the fourth-order perturbation are illustrated in Figs.~\ref{fig:sig3} and \ref{fig:sig4}.
By numerically calculating the coefficient of each order term, $\mathit{\Sigma}^{(2)}(k)$, $\mathit{\Sigma}^{(3)}(k)$ and $\mathit{\Sigma}^{(4)}(k)$, we investigate the convergence of the perturbation expansion in $U$.
Each term is evaluated by the following set of equations.

The bubble and ladder terms included in several diagrams are given by
\begin{subequations}
\begin{align}
\chi_0(q)&=-\sum_k \mathcal{G}_0(k)\mathcal{G}_0(k-q), \\[-2mm]
\phi_0(q)&=\hphantom{-}\sum_k \mathcal{G}_0(k)\mathcal{G}_0(q-k),
\end{align}
\end{subequations}
where $q=({\mib q},\mathrm{i}\nu_n)$, and $\nu_n=2n\pi T$ is the boson Matsubara frequency.
With these terms, the second- and the third-order terms are represented as
\begin{subequations}
\begin{align}
\mathit{\Sigma}^{(2)}(k)&=\hphantom{-}\sum_q \chi_0(q)\mathcal{G}_0(k-q), \\[-2mm]
             &=-\sum_q \phi_0(q)\mathcal{G}_0(q-k), \hspace{10mm}
\end{align}
\end{subequations}
and
\begin{equation}
\begin{split}
\mathit{\Sigma}^{(3)}(k)=\sum_q & \Bigl\{\chi_0(q)^2\mathcal{G}_0(k-q) \\[-5mm]
& \hspace{18mm} +\phi_0(q)^2\mathcal{G}_0(q-k)\Bigr\}.
\end{split}
\end{equation}
On the other hand, twelve terms in the fourth-order perturbation are classified into four groups. 
Each group contains three terms, which are exactly equivalent at half-filling $n=1$, (See Appendix)
\begin{equation}
\mathit{\Sigma}^{(4)}(k)=\mathit{\Sigma}^{(4)}_\mathrm{self}(k)+\mathit{\Sigma}^{(4)}_\mathrm{RPA}(k)
             +\mathit{\Sigma}^{(4)}_\mathrm{vtx1}(k)+\mathit{\Sigma}^{(4)}_\mathrm{vtx2}(k).
\end{equation}
Here, $\mathit{\Sigma}^{(4)}_\mathrm{self}(k)$ is the reducible self-energy including Green's function with the second-order self-energy.
The corresponding diagrams are denoted by $(a)-(c)$ in Fig.~\ref{fig:sig4}.
$\mathit{\Sigma}^{(4)}_\mathrm{RPA}(k)$ is the self-energy mediated by one boson fluctuation, which is represented by $(d)-(f)$ in Fig.~\ref{fig:sig4}.
These diagrams, which are included in the FLEX, display fluctuations of the longitudinal mode, transverse mode and ladder diagram, respectively.
$\mathit{\Sigma}^{(4)}_\mathrm{vtx1}(k)$ is the vertex correction for one boson fluctuation itself (type I), which is denoted by $(g)-(i)$ in Fig.~\ref{fig:sig4}.
These include, respectively, the longitudinal mode revised by itself, the ladder diagram revised by the transverse mode, and the transverse mode revised by the ladder diagram.
$\mathit{\Sigma}^{(4)}_\mathrm{vtx2}(k)$ is the vertex correction with two boson fluctuations crossing (type II), which is denoted by $(j)-(l)$ in Fig.~\ref{fig:sig4}.
$(j)$ denotes the longitudinal mode crossing with itself, and both $(k)$ and $(l)$ express the transverse mode crossing with the ladder diagram.
Each term is represented as
\begin{subequations}
\begin{align}
\begin{split}
\mathit{\Sigma}^{(4)}_\mathrm{self}(k)~=\sum_q 
& \Bigl\{2\chi_0(q)\mathcal{G}_1(k-q) \\[-5mm]
& \hspace{20mm} -\phi_0(q)\mathcal{G}_1(q-k)\Bigr\},
\end{split}
\label{eq:4thSelf}
\\
\begin{split}
\mathit{\Sigma}^{(4)}_\mathrm{RPA}(k)=\sum_q 
& \Bigl\{2\chi_0(q)^3\mathcal{G}_0(k-q) \\[-5mm]
& \hspace{18mm} -\phi_0(q)^3\mathcal{G}_0(q-k)\Bigr\},
\end{split}
\label{eq:4thRPA}
\\
\begin{split}
\mathit{\Sigma}^{(4)}_\mathrm{vtx1}(k)=\sum_q 
& \Bigl\{-\bigl(\chi_1(q)+\chi'_1(q)\bigr)\mathcal{G}_0(k-q) \\[-5mm]
& \hspace{20mm} +\phi_1(q)\mathcal{G}_0(q-k)\Bigr\},
\end{split}
\\
\begin{split}
\mathit{\Sigma}^{(4)}_\mathrm{vtx2}(k)=\sum_q 
& \Bigl\{\Lambda(k,q)\chi_0(q)\mathcal{G}_0(k-q) \\[-5mm]
& \hspace{7mm} +2\Lambda'(k,q)\phi_0(q)\mathcal{G}_0(q-k) \Bigr\},
\end{split}
\end{align}
\end{subequations}
where
\begin{equation}
\hspace{-22mm}\mathcal{G}_1(k)=\mathcal{G}_0(k)\mathit{\Sigma}^{(2)}(k)\mathcal{G}_0(k),
\end{equation}

\begin{subequations}
\begin{align}
\chi_1(q)&=\sum_{p}\Lambda(p,q)\mathcal{G}_0(p)\mathcal{G}_0(p-q), \\[-2mm]
\phi_1(q)&=\sum_{p}\Lambda'(p,q)\mathcal{G}_0(p)\mathcal{G}_0(q-p), \\[-2mm]
\chi'_1(q)&=\sum_{p}\Lambda''(p,q)\mathcal{G}_0(-p)\mathcal{G}_0(q-p),
\end{align}
\end{subequations}

\begin{subequations}
\begin{align}
\Lambda(k,q)&=\sum_{k'} \chi_0(k'-k)\mathcal{G}_0(k')\mathcal{G}_0(k'-q), \\[-2mm]
\Lambda'(k,q)&=\sum_{k'} \chi_0(k'-k)\mathcal{G}_0(k')\mathcal{G}_0(q-k'), \\[-2mm]
\Lambda''(k,q)&=\sum_{k'} \phi_0(k'-k)\mathcal{G}_0(k')\mathcal{G}_0(k'-q).
\end{align}
\end{subequations}
By computing all these terms, we examine the perturbation expansion to the fourth order.
However, it is very hard to calculate these terms, particularly, the vertex correction terms.
We need some technical procedures.
Before we proceed to numerical results, let us introduce them.

\section{Numerical Recipes}
In this section, we introduce some techniques of numerical calculations in the perturbation expansion. 
Fortunately, most of the terms in the self-energy possess the convolution form.
They can be easily evaluated with the use of the Fast Fourier Transform (FFT) algorithm. 
However, since $\mathcal{G}_0({\mib k},\mathrm{i}\omega_n)$ is not periodic as a function of $\omega_n$, applying the FFT to the Matsubara frequency sum yields extra numerical errors.
Instead, we start numerical calculations by $\mathcal{G}_0({\mib k},\tau)$, which is defined by
\begin{subequations}
\begin{align}
\mathcal{G}_0({\mib k},\tau)&=-\bigl(1-f(\xi_{\mib k})\bigr)\mathrm{e}^{-\xi_{\mib k}\tau} \\
&=-f(-|\xi_{\mib k}|)
\begin{cases}
\mathrm{e}^{\xi_{\mib k}(\beta-\tau)} &(\xi_{\mib k}<0) \\
\mathrm{e}^{-\xi_{\mib k}\tau}        &(\xi_{\mib k}>0).
\end{cases}
\end{align}
\end{subequations}
The final expression is convenient in numerical calculations due to few errors.
The Fourier transform is defined by
\begin{equation}
\mathcal{G}_0({\mib r},\tau)=\frac{1}{N^2}\sum_{\mib k} \mathcal{G}_0({\mib k},\tau)\mathrm{e}^{-\mathrm{i}{\mib k}\cdot{\mib r}}.
\end{equation}
With this relation, the bubble and ladder terms are expressed in the forms
\begin{subequations}
\label{eq:ch0}
\begin{align}
\chi_0({\mib q},\mathrm{i}\nu_n)
&=\sum_{\mib r}\mathrm{i}nt_0^\beta d\tau\chi_0({\mib r},\tau)\mathrm{e}^{\mathrm{i} qr}, \\[-2mm]
&=\sum_{\mib r}\mathrm{i}nt_0^\beta d\tau\mathcal{G}_0({\mib r},\tau)\mathcal{G}_0({\mib r},\beta-\tau)\mathrm{e}^{\mathrm{i} qr}, \\[-2mm]
\phi_0({\mib q},\mathrm{i}\nu_n)
&=\sum_{\mib r}\mathrm{i}nt_0^\beta d\tau\phi_0({\mib r},\tau)\mathrm{e}^{\mathrm{i} qr}, \\[-2mm]
&=\sum_{\mib r}\mathrm{i}nt_0^\beta d\tau\mathcal{G}_0({\mib r},\tau)^2\mathrm{e}^{\mathrm{i} qr},
\end{align}
\end{subequations}
where
\begin{subequations}
\begin{align*}
\mathrm{e}^{\mathrm{i} qr}&=\mathrm{e}^{\mathrm{i}{\mib q}\cdot{\mib r}}\mathrm{e}^{\mathrm{i}\nu_n\tau}, \\[-2mm]
\mathrm{i}nt_0^\beta d\tau \cdots &\simeq \frac{\beta}{N_\tau}\sum_\tau \cdots.
\end{align*}
\end{subequations}
Here, $N_\tau$ is the number of meshes along the imaginary time axis.
In this case, the cutoff of Matsubara frequencies is $\pi T N_\tau$.
In order to reduce numerical errors, we need to set this cutoff value large enough.
Although we can easily transform ${\mib r}$ into ${\mib q}$ in the above equations, we require some care in the integral over $\tau$.
$\chi_0({\mib q},\tau)$ and $\phi_0({\mib q},\tau)$ abruptly decrease far from $\tau=0$ or $\beta$.
In addition, $\phi_0({\mib q},\tau)$ has two different limiting values at $\tau=\pm 0$ or $\beta\pm 0$, and then does not strictly match with the FFT algorithm.
In order to avoid this problem, we use interpolation at around $\tau=0$ and $\beta$.
For a given ${\mib q}$, we carry out fitting by the function $f(\tau)=(a_0+a_1\tau)\exp(-a_2\tau)+(b_0+b_1\tau')\exp(-b_2\tau')$, with $\tau'=\beta-\tau$.
Then, $\delta\phi_0({\mib q},\tau)=\phi_0({\mib q},\tau)-f(\tau)$ is a smooth function at $\tau=0$ and $\beta$, where its value and slope almost vanish.
We can apply the FFT with high precision.
$f(\tau)$ itself can be easily integrated analytically.
Because of this careful treatment, $\chi_0({\mib q},\mathrm{i}\nu_n)$ and $\phi_0({\mib q},\mathrm{i}\nu_n)$ recover suitable $\nu_n$ dependences in the high-frequency region, and in the particle-hole symmetric case, satisfy the exact relation $\chi_0({\mib q}+{\mib Q},\mathrm{i}\nu_n)=\phi_0({\mib q},\mathrm{i}\nu_n)$ within numerical errors, where ${\mib Q}=(\pi,\pi)$. (See Appendix)
\begin{figure}[ht]
\begin{center}
\includegraphics[height=6cm]{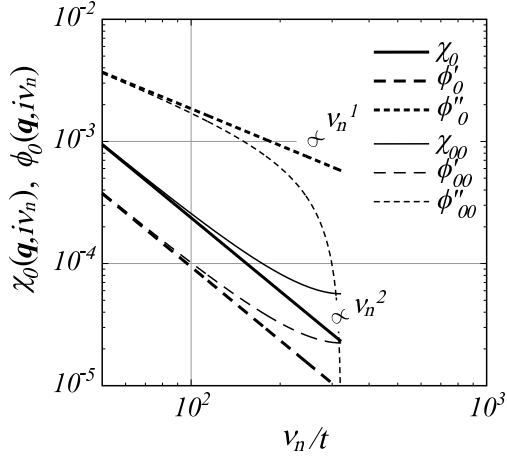}
\end{center}
\caption{$\chi_0({\mib q},\mathrm{i}\nu_n)$ and $\phi_0({\mib q},\mathrm{i}\nu_n)$ at ${\mib q}=(\pi,\pi/2)$ in the high-frequency region.
$\phi'_0$ and $\phi''_0$ represent the real and imaginary parts of $\phi_0$, respectively.
A subscript $0$ ($00$) corresponds to the calculation with (without) the interpolation.
$\chi_0$ and $\phi_0$ show suitable $\nu_n$ dependences in the high-frequency region, while $\chi_{00}$ and $\phi_{00}$ are curved owing to errors caused by applying the FFT.
This is the case of generic ${\mib q}$.
The numerical calculation is carried out for $(T,\mu)=(0.1t,-0.4t)$ with $N\times N\times N_\tau=64 \times 64 \times 1024$.}
\label{fig:chnu}
\end{figure}
In Fig.~\ref{fig:chnu}, we illustrate the $\nu_n$ dependences of $\chi_0({\mib q},\mathrm{i}\nu_n)$ and $\phi_0({\mib q},\mathrm{i}\nu_n)$ at ${\mib q}=(\pi,\pi/2)$ for $(T,\mu)=(0.1t,-0.4t)$ on the logarithmic scale as a sample.
One can see that the real and imaginary parts of them have suitable behaviors in the high-frequency region (respectively, $\nu_n^2$ and $\nu_n^1$ dependences), owing to the interpolation at around $\tau=0$ and $\beta$.
This is the case for generic ${\mib q}$, although $\chi_0({\mib q},\mathrm{i}\nu)$ at ${\mib q}=(0,0)$ vanishes for $\nu_n\ne 0$.

We next proceed to the calculation of the self-energy. 
The second- and third-order terms have a convolution form and are evaluated as
\begin{subequations}
\begin{align}
\label{eq:sig2}
&\mathit{\Sigma}^{(2)}(k)=\sum_r \mathcal{G}_0({\mib r},\tau)^2\mathcal{G}_0({\mib r},\beta-\tau)\mathrm{e}^{\mathrm{i} kr}, \\
\begin{split}
& \mathit{\Sigma}^{(3)}(k)=\sum_r \Bigl\{X_0({\mib r},\tau)\mathcal{G}_0(r) \\[-5mm]
& \hspace{28mm} -\Phi_0(r)\mathcal{G}_0({\mib r},\beta-\tau)\Bigr\}\mathrm{e}^{\mathrm{i} kr},
\end{split}
\label{eq:sig3}
\end{align}
\end{subequations}
where 
\begin{subequations}
\begin{align}
& X_0(r)   =\sum_q \chi_0(q)^2\mathrm{e}^{-\mathrm{i} qr}, \\[-2mm]
& \Phi_0(r)=\sum_q \phi_0(q)^2\mathrm{e}^{-\mathrm{i} qr}.
\end{align}
\end{subequations}
Here, the sum over $r=({\mib r},\tau)$ denotes
\begin{equation*}
\sum_r=\sum_{\mib r}\mathrm{i}nt_0^\beta d\tau\simeq \frac{\beta}{N_\tau}\sum_{\mib r}\sum_\tau.
\end{equation*}
In the particle-hole symmetric case, the third-order term vanishes within numerical errors owing to the interpolation, since the first term in eq.~(\ref{eq:sig3}) cancels out the second term.
Also in the fourth-order terms, the self-energy correction term eq.~(\ref{eq:4thSelf}) and the RPA term eq.~(\ref{eq:4thRPA}) possess the convolution form, and then are represented as 
\begin{equation}
\begin{split}
& \mathit{\Sigma}^{(4)}_\mathrm{self}(k)=\sum_r \Bigl\{2\chi_0(r)\mathcal{G}_1({\mib r},\tau) \\[-5mm]
& \hspace{32mm}+\phi_0(r)\mathcal{G}_1({\mib r},\beta-\tau)\Bigr\}\mathrm{e}^{\mathrm{i} kr},
\end{split}
\end{equation}
\begin{equation}
\begin{split}
& \mathit{\Sigma}^{(4)}_\mathrm{RPA}(k)=\sum_r \Bigl\{2X_1(r)\mathcal{G}_0({\mib r},\tau) \\[-5mm]
& \hspace{32mm} +\Phi_1(r)\mathcal{G}_0({\mib r},\beta-\tau)\Bigr\}\mathrm{e}^{\mathrm{i} kr},
\end{split}
\end{equation}
where 
\begin{subequations}
\begin{align}
& X_1(r)   =\sum_q \chi_0(q)^3\mathrm{e}^{-\mathrm{i} qr}, \\[-2mm]
& \Phi_1(r)=\sum_q \phi_0(q)^3\mathrm{e}^{-\mathrm{i} qr}.
\end{align}
\end{subequations}
All these terms in the convolution form can be calculated with the FFT algorithm.
However, the vertex correction terms $\mathit{\Sigma}^{(4)}_\mathrm{vtx1}(k)$ and $\mathit{\Sigma}^{(4)}_\mathrm{vtx2}(k)$ cannot be written in the convolution form.
Their evaluation requires very difficult computational effort.
A typical form in the vertex correction terms is that of $\chi_1(q)$, which included in $\mathit{\Sigma}^{(4)}_\mathrm{vtx1}(k)$.
We need an efficient way of computing it. 
Let us next introduce the technique adopted in this paper.

First of all, by the Fourier transform of $\chi_0(p_1-p_2)$, we decomposes $\chi_1(q)$ into the sum of the convolution form
\begin{align}
\begin{split}
& \chi_1(q)=\sum_{p_1,p_2} \Bigl\{\chi_0(p_1-p_2) \\[-5mm]
& \hspace{25mm} \times\mathcal{G}_0(p_1)\mathcal{G}_0(p_1+q)\mathcal{G}_0(p_2)\mathcal{G}_0(p_2+q)\Bigr\} 
\end{split}
\nonumber \\
\begin{split}
& \hspace{2mm} =\sum_{p_1,p_2}\sum_r \Bigl\{\chi_0(r)\mathrm{e}^{\mathrm{i}(p_1-p_2)r} \\[-5mm]
& \hspace{25mm} \times\mathcal{G}_0(p_1)\mathcal{G}_0(p_1+q)\mathcal{G}_0(p_2)\mathcal{G}_0(p_2+q) \Bigr\} 
\end{split}
\nonumber \\
\begin{split}
& \hspace{2mm}=\sum_r \chi_0(r)
  \Bigl\{ \sum_{p_1}\mathrm{e}^{\mathrm{i} p_1r}\mathcal{G}_0(p_1)\mathcal{G}_0(p_1+q) \Bigr\} \\[-3mm]
& \hspace{25mm} 
  \times\Bigl\{ \sum_{p_2}\mathrm{e}^{-\mathrm{i} p_2r}\mathcal{G}_0(p_2)\mathcal{G}_0(p_2+q) \Bigr\}.
\end{split}
\end{align}
In the final expression, for a given $r$, the sum over $p_1$ and $p_2$ can be evaluated with the FFT algorithm. 
However, the calculation time, which rapidly increases with a square of $M=N^2N_\tau$, does not change yet.
It is almost impossible to do it within a practical machine time. 
Fortunately, we can restrict the sum over ${\mib r}$ to a number much less than $N^2$, considering a characteristic property of $\chi_0({\mib r},\tau)$. 
We next demonstrate it.

In order to obtain the proper perspective and reduce numerical errors, we classify the sum over ${\mib r}$ with the use of the space group $C_{4v}$ symmetry. 
$\chi_0({\mib r},\tau)$ has the same value at any ${\mib r}$ with the same distance $|{\mib r}|$.
There are eight equivalent points in generic ${\mib r}$.
(four equivalent points on the $x$- and $y$-axis and the boundary, and so on.)
The sum of eight exponential functions with the same distance $|{\mib r}|$ are transformed into the sum of eight $C_{4v}$ basis functions expressed as
\begin{equation}
\sum_{{\mib r}}\mathrm{e}^{\mathrm{i}({\mib p}_1-{\mib p}_2)\cdot{\mib r}}=\sum_{|{\mib r}|,n}\phi_n({\mib p}_1)\phi_n({\mib p}_2),
\end{equation}
where eight $C_{4v}$ basis functions $\phi_n({\mib p})$ at generic ${\mib r}=(i,j)$ are classified with irreducible representations (IR), as shown in Table.\ref{table:IR}.
\begin{table}[ht]
\caption{$C_{4v}$ basis functions $\phi_n({\mib p})$ at ${\mib r}=(i,j)$}
\label{table:IR}
\begin{center}
\begin{tabular}{clc}
\hline 
$n$ & IR & Basis functions \\
\hline
1&$A_1$ & $\displaystyle \sqrt{2}\bigl(\cos(ip_x)\cos(jp_y)+\cos(jp_x)\cos(ip_y)\bigr)$ \\
2&$A_2$ & $\displaystyle \sqrt{2}\bigl(\sin(ip_x)\sin(jp_y)-\sin(jp_x)\sin(ip_y)\bigr)$ \\
3&$B_1$ & $\displaystyle \sqrt{2}\bigl(\cos(ip_x)\cos(jp_y)-\cos(jp_x)\cos(ip_y)\bigr)$ \\
4&$B_2$ & $\displaystyle \sqrt{2}\bigl(\sin(ip_x)\sin(jp_y)+\sin(jp_x)\sin(ip_y)\bigr)$ \\
5&$E$   & $\displaystyle 2\sin(ip_x)\cos(jp_y)$ \\
6&$E$   & $\displaystyle 2\sin(ip_y)\cos(jp_x)$ \\
7&$E$   & $\displaystyle 2\sin(jp_x)\cos(ip_y)$ \\
8&$E$   & $\displaystyle 2\sin(jp_y)\cos(ip_x)$ \\
\hline
\end{tabular}
\end{center}
\end{table}
Precisely speaking, the subscript $n$ in $\phi_n({\mib p})$ should be $|{\mib r}|$ and $n$, where $n$ denotes a kind of IR.
To simplify the expression, however, we do not explicitly write $|{\mib r}|$ as a subscript of functions, hereafter.
Instead, to explicitly show the expansion in $C_{4v}$ basis functions, we denote $\chi_0({\mib r},\tau)$ by $\chi^0_{n\tau}$.
Thus,
\begin{align}
\begin{split}
& \chi_1(q)=\sum_{|{\mib r}|n\tau} \chi^0_{n\tau}
  \Bigl\{ \sum_{p_1}\phi_n({\mib p}_1)\mathcal{G}_0(p_1)\mathcal{G}_0(p_1+q)\mathrm{e}^{ \mathrm{i}\omega_1\tau} 
  \Bigr\} \\[-3mm]
& \hspace{20mm} \times
  \Bigl\{ \sum_{p_2}\phi_n({\mib p}_2)\mathcal{G}_0(p_2)\mathcal{G}_0(p_2+q)\mathrm{e}^{-\mathrm{i}\omega_2\tau}
  \Bigr\},
\end{split}
\nonumber \\
& \hspace{8mm}=\sum_{|{\mib r}|n\tau}\chi^0_{n\tau}f_{n\tau}(q)f_{n-\tau}(q),
\label{eq:ch1}
\end{align}
where 
\begin{subequations}
\begin{align}
\begin{split}
& f_{n\tau}(q)=\sum_{p_1}
\phi_n({\mib p}_1)\mathcal{G}_0(p_1)\mathcal{G}_0(p_1+q)\mathrm{e}^{\mathrm{i}(\omega_1+\nu)\tau} \\[-2mm]
&\hspace{10mm}=\sum_{p_1}
\bar{\mathcal{G}}_n(p_1)\mathcal{G}_0(p_1+q)\mathrm{e}^{\mathrm{i}(\omega_1+\nu)\tau}, \\[-2mm]
\end{split}
\\
\begin{split}
& \bar{\mathcal{G}}_n(p_1)=\phi_n({\mib p}_1)\mathcal{G}_0(p_1).
\end{split}
\end{align}
\end{subequations}
Since $f_{n\tau}(q)$ is in the convolution form, it is evaluated by calculating the Fourier transform as
\begin{equation}
f_{n\tau}(q)=\sum_{r_1}
\bar{\mathcal{G}}_n({\mib r}_1,\tau_1)\mathcal{G}_0({\mib r}_1,-\tau_1-\tau)\mathrm{e}^{-\mathrm{i} qr_1}.
\end{equation}
As will be shown later, we take at most 20 points with large contributions for the sum over $|{\mib r}|$ in eq.~(\ref{eq:ch1}). 
This can be carried out using recent supercomputers. 
In this way, we can compute the vertex correction terms $\mathit{\Sigma}^{(4)}_\mathrm{vtx1}(k)$ and $\mathit{\Sigma}^{(4)}_\mathrm{vtx2}(k)$. 
Finally, the type-I vertex correction term is given by 
\begin{equation}
\begin{split}
\mathit{\Sigma}^{(4)}_\mathrm{vtx1}(k)=\sum_{r} &
\Bigl\{-\bigl(X_2(r)+X'_2(r)\bigr)\mathcal{G}_0({\mib r},\tau)\mathrm{e}^{\mathrm{i} kr} \\[-5mm]
& \hspace{14mm} -\Phi_2(r)\mathcal{G}_0({\mib r},\beta-\tau)\mathrm{e}^{\mathrm{i} kr}\Bigr\}
\end{split}
\end{equation}
where 
\begin{subequations}
\begin{align}
X_2(r)   =\sum_q \chi_1(q)\mathrm{e}^{-\mathrm{i} qr}, \\[-2mm]
X'_2(r)  =\sum_q \chi'_1(q)\mathrm{e}^{-\mathrm{i} qr}, \\[-2mm]
\Phi_2(r)=\sum_q \phi_1(q)\mathrm{e}^{-\mathrm{i} qr}.
\end{align}
\end{subequations}
\begin{subequations}
\begin{align}
& \chi_1(q) =\sum_{|{\mib r}|n\tau} 
\chi^0_{n\tau}f_{n\tau}(q)f_{n-\tau}(q), \\[-2mm]
& \chi'_1(q)=\sum_{|{\mib r}|n\tau} 
\phi^0_{n\tau}f_{n\tau}(q)f_{n\tau}(q)^*, \\[-2mm]
& \phi_1(q) =\sum_{|{\mib r}|n\tau} \chi^0_{n\tau}f'_{n\tau}(q)f'_{n-\tau}(q),
\end{align}
\end{subequations}
\begin{equation}
\begin{split}
f'_{n\tau}(q)&=\sum_{p_1}
\bar{\mathcal{G}}_n(p_1)\mathcal{G}_0(q-p_1)\mathrm{e}^{\mathrm{i}(\nu-\omega_1)\tau} \\[-2mm]
&=\sum_{r_1}\bar{\mathcal{G}}_n({\mib r}_1,\tau_1)\mathcal{G}_0({\mib r}_1,\tau_1-\tau)\mathrm{e}^{\mathrm{i} qr_1}.
\end{split}
\end{equation}
The type-II vertex correction term is given by
\begin{align}
\begin{split}
& \mathit{\Sigma}^{(4)}_\mathrm{vtx2}=\sum_{|{\mib r}|n\tau}\sum_{k',q} 
\Bigl\{\chi^0_{n\tau}\bar{\mathcal{G}}_n(k')\mathcal{G}_0(k'+q)\mathrm{e}^{\mathrm{i}(\omega'+\nu)\tau} \\[-5mm]
& \hspace{42mm} \times\chi_0(q)\bar{\mathcal{G}}_n(k-q)\mathrm{e}^{-\mathrm{i}\omega\tau}
\end{split}
\nonumber \\
\begin{split}
& \hspace{20mm}
+2\chi^0_{n\tau}\bar{\mathcal{G}}_n(k')\mathcal{G}_0(q-k')\mathrm{e}^{\mathrm{i}(\nu-\omega')\tau} \\[-2mm]
& \hspace{42mm} \times\phi_0(q)\bar{\mathcal{G}}_n(q-k)\mathrm{e}^{-\mathrm{i}\omega\tau} \Bigr\}
\end{split}
\nonumber \\
\begin{split}
&=\sum_{|{\mib r}|n\tau}\sum_q \chi^0_{n\tau}\mathrm{e}^{-\mathrm{i}\omega\tau}\Bigl\{
f_{n\tau}(q)\chi_0(q)\bar{\mathcal{G}}_n(k-q) \\[-5mm]
& \hspace{38mm}+2f'_{n\tau}(q)\phi_0(q)\bar{\mathcal{G}}_n(q-k)\Bigr\}
\end{split}
\nonumber \\
& =\sum_{|{\mib r}|n\tau}\sum_{r_1} 
\chi^0_{n\tau}\mathrm{e}^{-\mathrm{i}\omega\tau}\mathrm{e}^{\mathrm{i} kr_1} \\[-2mm]
& \hspace{5mm} \times\Bigl\{\tilde f_{n\tau}(r_1)\tilde{\mathcal{G}}_n({\mib r}_1,\tau_1) 
-2\tilde f'_{n\tau}(r_1)\tilde{\mathcal{G}}_n({\mib r}_1,\beta-\tau_1)\Bigr\}.
\nonumber
\end{align}
When $\phi_n(p)$ is an even function ($n=1,2,3,4$),
\begin{subequations}
\begin{align}
\tilde{\mathcal{G}}_n(r)&=\sum_k \bar{\mathcal{G}}_n(k)\mathrm{e}^{-\mathrm{i} kr}, \\[-2mm]
\tilde f_{n\tau}(r)&=\sum_q f_{n\tau}(q)\chi_0(q)\mathrm{e}^{-\mathrm{i} qr}, \\[-2mm]
\tilde f'_{n\tau}(r)&=\sum_q f'_{n\tau}(q)\phi_0(q)\mathrm{e}^{-\mathrm{i} qr},
\end{align}
\end{subequations}
and when $\phi_n(p)$ is an odd function ($n=5,6,7,8$),
\begin{subequations}
\begin{align}
\tilde{\mathcal{G}}_n(r)
&  =-\mathrm{i}\sum_k \bar{\mathcal{G}}_n(k)\mathrm{e}^{-\mathrm{i} kr}, \\[-2mm]
\tilde f_{n\tau}(r)
&  =\hphantom{+}\mathrm{i}\sum_q f_{n\tau}(q)\chi_0(q)\mathrm{e}^{-\mathrm{i} qr}, \\[-2mm]
\tilde f'_{n\tau}(r)
&  =-\mathrm{i}\sum_q f'_{n\tau}(q)\phi_0(q)\mathrm{e}^{-\mathrm{i} qr}.
\end{align}
\end{subequations}
Thus, we can calculate a set of equations for the self-energy in the fourth-order perturbation within a practical machine time.
\begin{figure}[t!]
\begin{center}
\includegraphics[height=5cm]{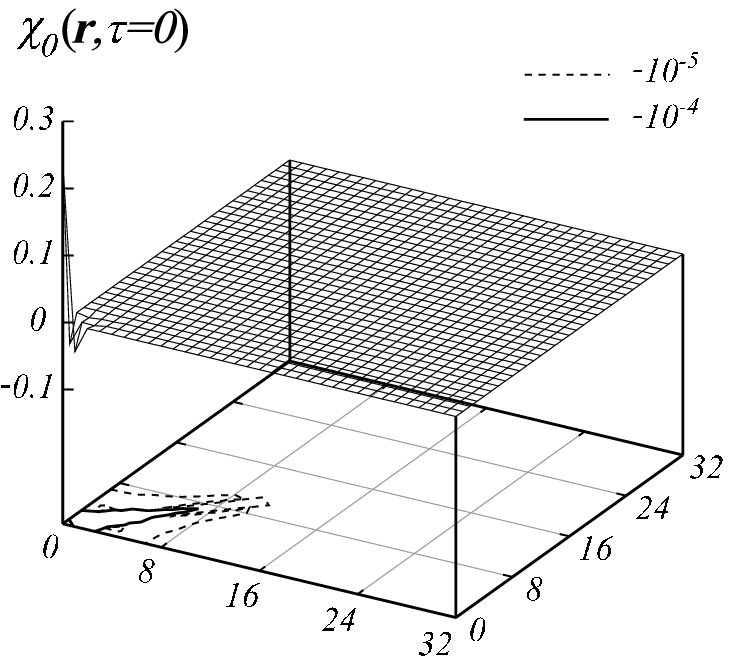} \\
\vspace{10 pt}
\includegraphics[height=5cm]{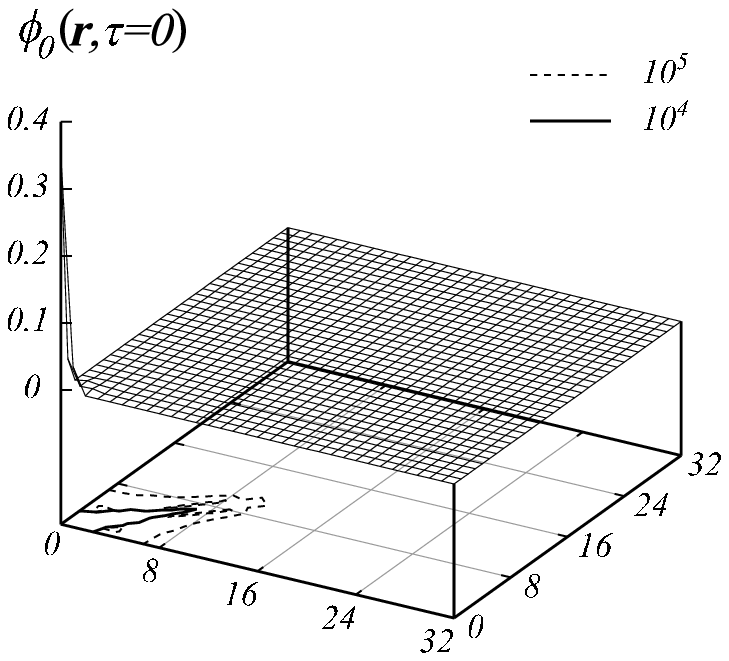}
\caption{$\chi_0({\mib r},\tau=0)$ and $\phi_0({\mib r},\tau=0)$ for $(T,\mu)=(0.1t,-0.4t)$.
They have large magnitudes only at around ${\mib r}=(0,0)$.}
\label{fig:chr}
\end{center}
\end{figure}
\begin{figure}[t!]
\begin{center}
\includegraphics[height=5cm]{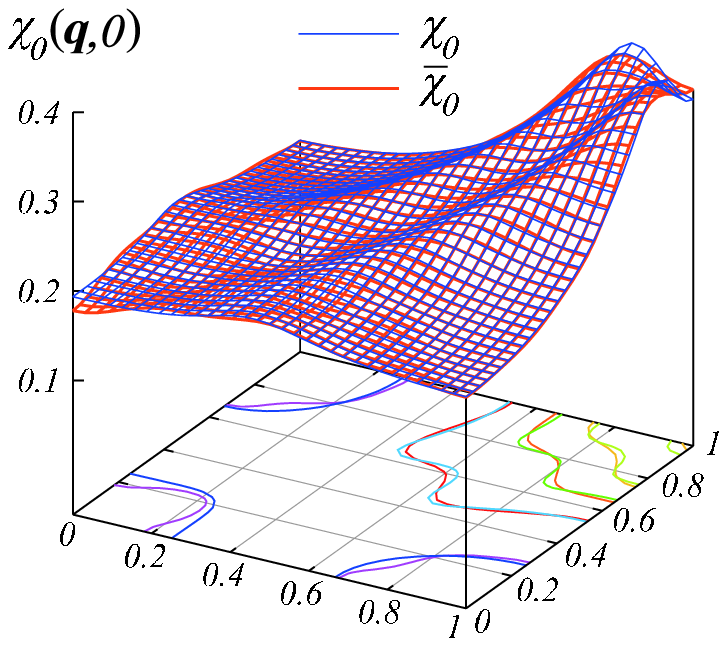} \\
\vspace{10 pt}
\includegraphics[height=5cm]{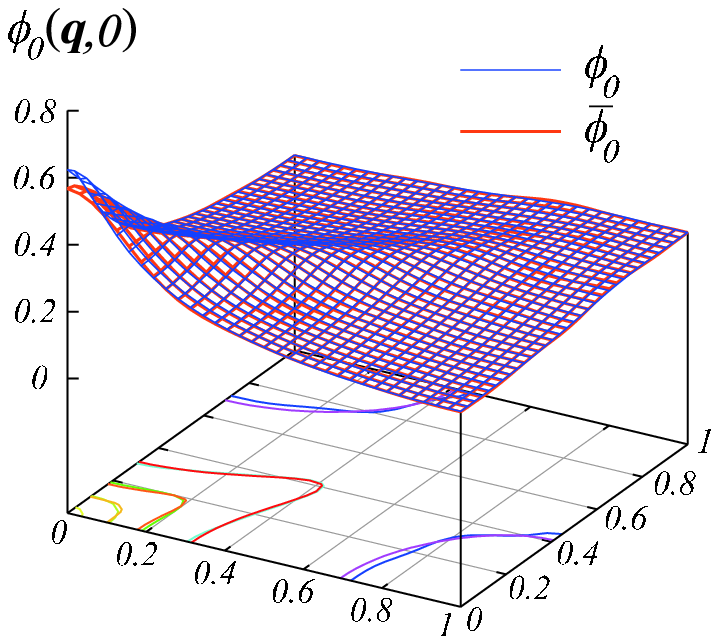}
\caption{(Color online) $\chi_0({\mib q},0)$ and $\phi_0({\mib q},0)$ calculated using eq.~(\ref{eq:ch0}), and $\bar\chi_0({\mib q},0)$ and $\bar\phi_0({\mib q},0)$ calculated using the restricted ${\mib r}$ summation for $(T,\mu)=(0.1t,-0.4t)$.
Although the structure of the latter functions $\bar\chi_0$ and $\bar\phi_0$ is somewhat smeared, they well restore the overall features.
The unit of the $x$- and $y$-axes is $\pi/a$, where $a$ is the lattice constant.}
\label{fig:chp}
\end{center}
\end{figure}

Before we finish this section, let us show that the restricted ${\mib r}$ summation is efficient. 
In Fig.~\ref{fig:chr}, we display $\chi_0({\mib r},\tau=0)$ and $\phi_0({\mib r},\tau=0)$. 
It is clear that they have large magnitudes at around ${\mib r}=(0,0)$. 
We can approximately represent the ${\mib r}$ summation by several points at around ${\mib r}=(0,0)$.
In this paper, we restrict the ${\mib r}$ summation to several points around ${\mib r}=(0,0)$ at which the magnitude of functions exceeds $10^{-4}$.
This corresponds to the fact that $\chi_0(q)$ and $\phi_0(q)$ are approximated by the smeared functions $\bar\chi_0(q)$ and $\bar\phi_0(q)$, as illustrated in Fig.~\ref{fig:chp}.
One can see that the smeared functions restore the overall features.
In this case, we can confirm the quality of the approximation by adopting the above-mentioned method for the second-order self-energy.
\begin{align}
\label{eq:sig2a}
\mathit{\Sigma}^{(2)}(k)
&=\sum_{k'}\chi_0(k-k')\mathcal{G}_0(k') \nonumber \\[-2mm]
&=\sum_{k'}\sum_{|{\mib r}|n\tau}\chi^0_{n\tau}\phi_n({\mib k})\phi_n({\mib k}')
  \mathrm{e}^{\mathrm{i}\omega\tau}\mathrm{e}^{-\mathrm{i}\omega'\tau}\mathcal{G}_0(k') \nonumber \\[-2mm]
&=\sum_{|{\mib r}|n\tau}\chi^0_{n\tau}\phi_n({\mib k})\mathrm{e}^{\mathrm{i}\omega\tau}
  \sum_{k'}\bar{\mathcal{G}}_n(k')\mathrm{e}^{-\mathrm{i}\omega'\tau} \nonumber \\[-2mm]
&=\sum_{|{\mib r}|n\tau}
  \phi_n({\mib k})\chi^0_{n\tau}\bar{\mathcal{G}}_n(0,\tau)\mathrm{e}^{\mathrm{i}\omega\tau} \nonumber \\[-2mm]
&=\sum_{|{\mib r}|}\phi_1({\mib k})
  \sum_\tau\chi^0_{1\tau}\bar{\mathcal{G}}_1(0,\tau)\mathrm{e}^{\mathrm{i}\omega\tau}.
\end{align}
The last equality results from the fact that $\bar{\mathcal{G}}_n(0,\tau)$ is finite,
only when $\phi_n({\mib k}')$ has $A_1$ symmetry (i.e., $n=1$).
\begin{figure}[t!]
\begin{center}
\includegraphics[height=65mm]{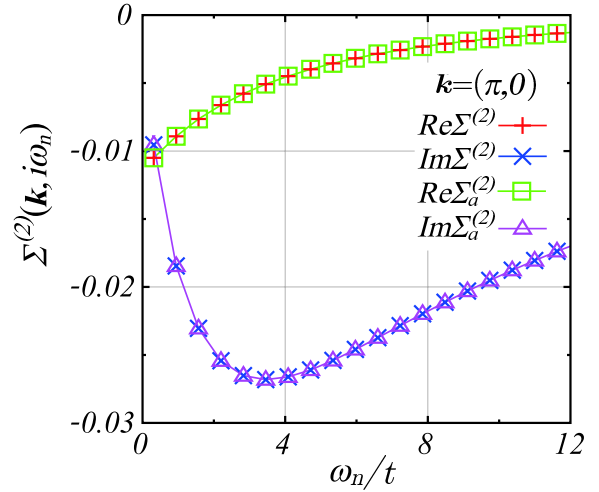}
\caption{(Color online) Real and imaginary parts of the second-order term at ${\mib k}=(\pi,0)$ for $(T,\mu)=(0.1t,-0.4t)$.
We can see a small difference in quantity between the direct calculation of the convolution form in eq.~(\ref{eq:sig2}), $\mathit{\Sigma}^{(2)}({\mib k},\mathrm{i}\omega_n)$, and the approximate value by the restricted ${\mib r}$ summation, $\mathit{\Sigma}^{(2)}_a({\mib k},\mathrm{i}\omega_n)$ in eq.~(\ref{eq:sig2a}).}
\label{fig:sigcheck}
\end{center}
\end{figure}
Figure \ref{fig:sigcheck} exhibits a comparison between the direct calculation of the convolution form $\mathit{\Sigma}^{(2)}(k)$ and the approximate value by the restricted ${\mib r}$ summation $\mathit{\Sigma}^{(2)}_a(k)$.
We can see little difference in quantity between these two self-energies on this scale.
The relative errors are $|\mathit{\Sigma}^{(2)}(k)-\mathit{\Sigma}^{(2)}_a(k)|/|\mathit{\Sigma}^{(2)}(k)|\lesssim 6\times 10^{-3}$.
Thus, this approximation proves to be efficient.
This fact can also be verified in the third-order self-energy.

\section{Numerical Results}
Using the technique introduced in \S 3, we calculate all terms up to the fourth order in self-energy.
In this paper, we carry out practical numerical calculations for $M=N^2N_\tau=64 \times 64 \times 1024$.
We set $T=0.1t$ except in the case of the discussion of temperature dependence.
First, we investigate the convergence of the perturbation expansion.
Next, we demonstrate the behavior of single-particle quantities, such as the DOS.

\subsection{Convergence of perturbation expansion}

As mentioned in \S 1, in the perturbation expansion for the impurity Anderson model, physical quantities behave like the exponential function as a function of $U$.
Thus, the coefficient of $U^n$ in the perturbation series rapidly decreases approximately in proportion to $\sim 1/n!$. 
Owing to this property, even if we truncate the perturbation expansion within a finite order, these physical quantities rapidly approach the exact values with increasing cutoff order.
In the impurity Anderson model, the fourth-order perturbation possesses a sufficient accuracy in a moderate correlation regime. 
Here, with these well-known facts in mind, let us examine the convergence of the perturbation expansion for the self-energy in the lattice system.
We first study the half-filled case, and next, the doped case, and then, discuss the validity of the perturbation expansion truncated within a finite order.

{\it Half-filled case} \rule[2pt]{24pt}{0.5pt}
At half-filling $n=1$, the chemical potential $\mu=0$.
In this case, the system possesses particle-hole symmetry.
The FS is just at the AF Brillouin zone boundary.
Owing to the complete nesting with ${\mib Q}=(\pi,\pi)$, the spin susceptibility $\chi({\mib Q},0)$ will diverge at the Neel temperature $T_\mathrm{N}$.
The system undergoes the phase transition to the AF phase, although $T_\mathrm{N} \rightarrow 0$ by the Mermin-Wargner theorem in exact two-dimensional systems.
The FL state becomes unstable in the ground state.
For $T>T_\mathrm{N}$, however, the system can stay in the FL state, if we consider Anderson's continuation principle.
In this section, we explain that the perturbation expansion can be valid for $T \gtrsim 0.1t$ at least, and then, for large $U\gtrsim 6t$, the FL state seems to break down partially.
The remarkable features become clearer by calculating single-particle quantities, such as the DOS.

\begin{figure}[t!]
\begin{center}
\includegraphics[height=65mm]{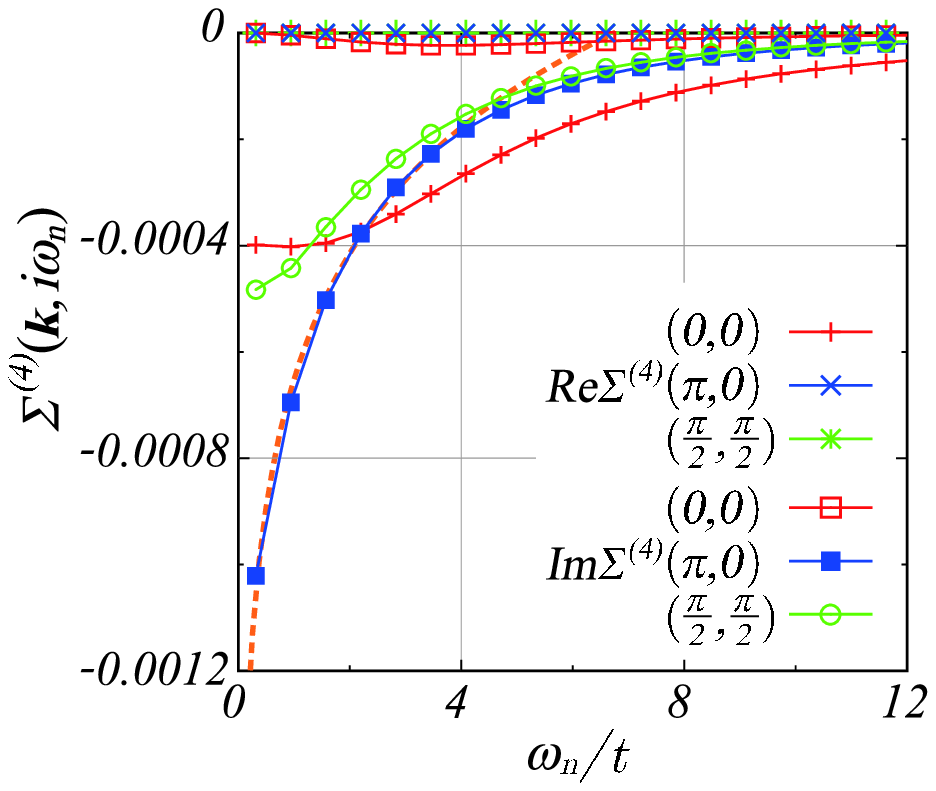}
\caption{(Color online) Real and imaginary parts of the fourth-order term $\mathit{\Sigma}^{(4)}({\mib k},\mathrm{i}\omega_n)$ at ${\mib k}=(0,0)$, $(\pi,0)$ and $(\pi/2,\pi/2)$ at $T=0.1t$ in the half-filled case.
The imaginary part at ${\mib k}=(\pi,0)$ in the low-frequency region provides the largest magnitude.
It is increasing roughly according to the logarithmic function, $0.00034\log(0.148\omega_n)$, represented by the dotted line.}
\label{fig:sig4thhalf}

\vspace{10 pt}
\includegraphics[height=65mm]{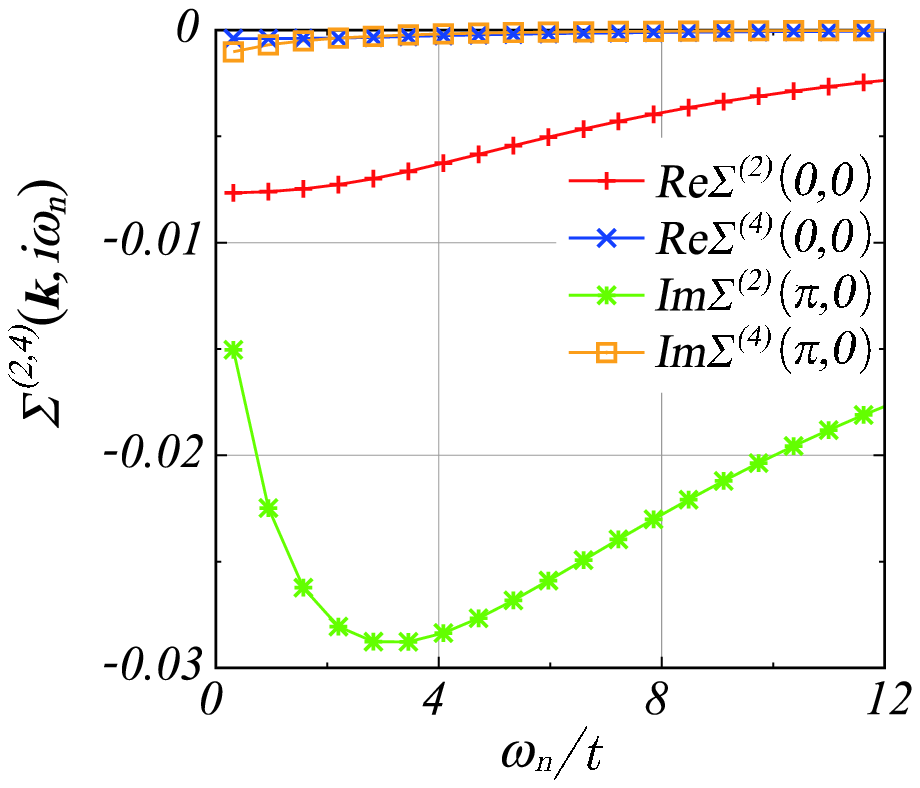}
\caption{(Color online) Comparison between the second-order term $\mathit{\Sigma}^{(2)}({\mib k},\mathrm{i}\omega_n)$ and the fourth-order term $\mathit{\Sigma}^{(4)}({\mib k},\mathrm{i}\omega_n)$ in the half-filled case.
We display the real parts at ${\mib k}=(0,0)$ and the imaginary parts at ${\mib k}=(\pi,0)$.
At a glance, one can see that $\mathit{\Sigma}^{(4)}({\mib k},\mathrm{i}\omega_n)$ is much smaller than $\mathit{\Sigma}^{(2)}({\mib k},\mathrm{i}\omega_n)$.}
\label{fig:sighalf}
\end{center}
\end{figure}
In Fig.~\ref{fig:sig4thhalf}, we display $\omega_n$ dependence of the fourth-order term $\mathit{\Sigma}^{(4)}({\mib k},\mathrm{i}\omega_n)$ at the two points on the FS, ${\mib k}=(\pi,0)$ and $(\pi/2,\pi/2)$, and at the $\mathit{\Gamma}ma$ point far from the FS, ${\mib k}=(0,0)$.
The real part $\mathrm{Re}\mathit{\Sigma}^{(4)}({\mib k},\mathrm{i}\omega_n)$ is equal to zero on the FS, and its magnitude is the largest at the $\mathit{\Gamma}ma$ point, while the imaginary part $\mathrm{Im}\mathit{\Sigma}^{(4)}({\mib k},i\omega_n)$ has the largest magnitude at ${\mib k}=(\pi,0)$.
The behavior of the real part on the FS indicates that the FS is not deformed by the electron correlation, since the real part at $\omega_n\to 0$ limit is connected with an energy shift after the analytic continuation on the real axis.
In Fig.~\ref{fig:sighalf}, we compare the largest value in the fourth-order term with that in the corresponding second-order term.
The third-order term vanishes due to particle-hole symmetry. (See Appendix)
The overall behavior of $\mathrm{Im}\mathit{\Sigma}^{(2)}({\mib k},i\omega_n)$ is almost independent of the wave number ${\mib k}$.
At a glance, one can find that the fourth-order term is much smaller than the second-order term.
This means that the perturbation expansion converges rapidly.
\begin{figure}[t!]
\begin{center}
\includegraphics[height=65mm]{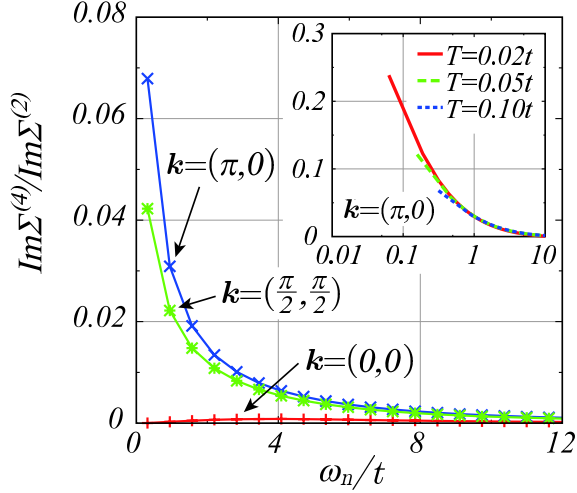}
\caption{(Color online) Ratio of $\mathrm{Im}\mathit{\Sigma}^{(4)}({\mib k},i\omega_n)$ to $\mathrm{Im}\mathit{\Sigma}^{(2)}({\mib k},i\omega_n)$ at ${\mib k}=(0,0)$, $(\pi,0)$ and $(\pi/2,\pi/2)$ in the half-filled case.
The ratio is the largest at ${\mib k}=(\pi,0)$ at any frequencies, and the maximum value is smaller than $\frac{1}{4!}/\frac{1}{2!}=1/12\simeq 0.08$.
In particular, the ratio is negligible in the high-frequency region.
The inset denotes the ratio at ${\mib k}=(\pi,0)$ for $T=0.02t$, $0.05t$ and $0.1t$ as a function of $\log(\omega_n/t)$.
The largest value at $\omega_n=\pi T$ quickly increases over $1/12\simeq 0.08$.}
\label{fig:ratio}
\end{center}
\end{figure}
In Fig.~\ref{fig:ratio}, we evaluate the ratio of the fourth-order term to the second-order term, quantitatively.
For simplicity, we display only the ratio of the imaginary part, because it is larger than the real part, and never crosses $0$.
In addition, the behavior is important for the behavior of the quasiparticle on the FS, since the imaginary part at $\omega_n\to 0$ limit is connected with the mass renormalization factor after the analytic continuation.
In Fig.~\ref{fig:ratio}, the ratio is large on the FS, and becomes the largest at ${\mib k}=(\pi,0)$.
Although the value increases with smaller $\omega_n$, the maximum value is $\sim 0.07$.
If we expect the same degree of convergence in the perturbation expansion as $\sim 1/n!$ in the impurity case, then the ratio has to be of the order of $\frac{1}{4!}/\frac{1}{2!}=1/12\simeq 0.08$.
Thus, in this case, the convergence of the perturbation expansion is considered good.
At lower temperatures, however, the maximum value becomes larger than $1/12$ as shown in the inset of Fig.~\ref{fig:ratio}.
Therefore, for $T \lesssim 0.1t$, the perturbation expansion may break, or at least we need higher-order terms.
In order to clarify this point, we need to evaluate the sixth-order term.
However, this is difficult to perform.
On the other hand, in the high-frequency region, the ratio is almost zero.
The convergence of the perturbation expansion is very good.
Thus, for the present case, we can expect that the perturbation expansion is applicable for $T\gtrsim 0.1t$, since the coefficient of each order term becomes rapidly small as in the impurity case.

\begin{figure}[t!]
\begin{center}
\includegraphics[height=65mm]{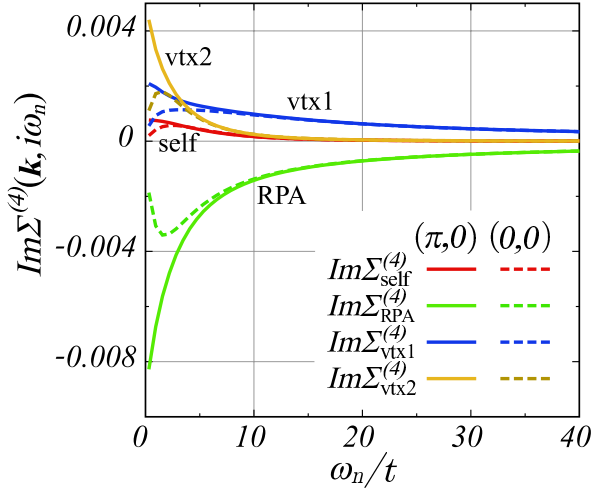}
\caption{(Color online) Imaginary part of each type of term in the fourth-order self-energy at ${\mib k}=(\pi,0)$ and $(0,0)$ in the half-filled case.
The large negative value of the RPA term is cancelled out by the other terms almost completely in the high-frequency region.}
\label{fig:sigpartshalf}
\end{center}
\end{figure}
Next, let us separately examine the contributions of each type of term in the fourth-order self-energy.
In Fig.~\ref{fig:sigpartshalf}, we show the imaginary part of each term at ${\mib k}=(\pi,0)$ and $(0,0)$.
Only the RPA term has a large negative value, and the other terms behave to compensate for it.
Although its magnitude at ${\mib k}=(\pi,0)$ continues to increase in the low-frequency region, it turns to decrease at generic ${\mib k}$ points far from the FS.
As a whole, the large value in the low-frequency region becomes one-order smaller due to the compensation.
On the other hand, in the high-frequency region, the self-energy correction term and the type-II vertex correction term are suppressed abruptly.
The RPA term and the type-I vertex correction term have a long tail with opposite signs.
These terms cancel out each other almost completely in the high-frequency region.
Therefore, the fourth-order term as a whole vanishes rapidly in the high-frequency region.
Thus, the behavior of self-energy in the high-frequency region is dominated by the second-order term.
This indicates that the simple perturbation expansion up to the fourth order keeps the exact atomic limit behavior $U^2/4\mathrm{i}\omega_n$ in the high-frequency region.~\cite{rf:Hewson}
This is different from the approximate method involving the resummation of the specific diagrams, such as the FLEX.
This fact is important in realizing the Mott-Hubbard character.
The failure of the incoherent Hubbard peak structure in the FLEX comes from this fact partially.

Let us consider the behavior in the low-frequency region in details.
The increase in the ratio at $(\pi,0)$ originates from the increase in the fourth-order term itself.
It roughly behaves like $\log(\omega_n)$ as shown in Fig.~\ref{fig:sig4thhalf}.
Such behavior is different from the conventional FL behavior.
Generally, the imaginary part of self-energy in the FL state has a tendency to decrease its magnitude in the low-frequency region such as that observed for the second-order term.
This implies that the self-energy includes a term proportional to $\mathrm{i}\omega_n$ in the expansion at around $\omega_n\to 0$.
This is equivalent to existence of the weight of the FL quasiparticle.
Thus, the singular behavior like $\log(\omega_n)$ in the fourth-order term is not the conventional FL behavior, and rather, is regarded as the precursor into the Mott transition, although the singularity is weaker than $1/\mathrm{i}\omega_n$ in the Mott transition.
\begin{figure}[t!]
\begin{center}
\includegraphics[height=65mm]{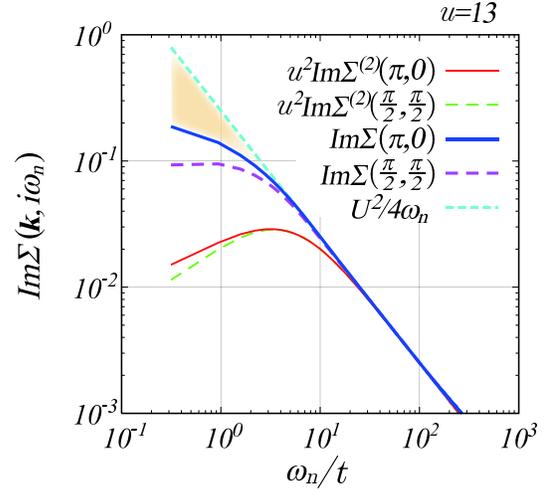}
\caption{(Color online) Magnitude of the imaginary part of the self-energy up to the second-order, $u^2\mathrm{Im}\mathit{\Sigma}^{(2)}(k)$, and up to the fourth order, $\mathrm{Im}\mathit{\Sigma}(k)=u^2\mathrm{Im}\mathit{\Sigma}^{(2)}(k)+u^4\mathrm{Im}\mathit{\Sigma}^{(4)}(k)$, at ${\mib k}=(\pi,0)$ and $(\pi/2,\pi/2)$ for $U=13t$ on the logarithmic scale.
The unit of the vertical axis is $u^2t=U^2/t$.
The self-energy to the fourth order is closer to the atomic limit behavior, $U^2/4\mathrm{i}\omega_n$.
We can expect that most of contributions from the higher-order terms are confined in the shaded area in the low-frequency region.}
\label{fig:Mott}
\end{center}
\end{figure}
In Fig.~\ref{fig:Mott}, we can clearly confirm the asymptotic behavior.
It indicates the magnitude of the imaginary part of the self-energy up to the second order, $u^2\mathrm{Im}\mathit{\Sigma}^{(2)}(k)$, and up to the fourth order, $\mathrm{Im}\mathit{\Sigma}(k)=u^2\mathrm{Im}\mathit{\Sigma}^{(2)}(k)+u^4\mathrm{Im}\mathit{\Sigma}^{(4)}(k)$, at ${\mib k}=(\pi,0)$ and $(\pi/2,\pi/2)$ for large $U=13t$.
They asymptotically approach the atomic limit $U^2/4\mathrm{i}\omega_n$ regardless of ${\mib k}$ in the high-frequency region.
For the large $U=13t$, the fourth-order perturbation is closer to the atomic limit form than the second-order perturbation.
This implies that the fourth-order perturbation can reflect the Mott-Hubbard character more strongly, which has not been grasped in the second-order perturbation so far.
In fact, as indicated later, the DOS shows the pseudogap behavior like the Mott-Hubbard AF gap, and the spectral weight at ${\mib k}=(\pi,0)$ does not indicate the quasiparticle peak at $\omega=0$ any longer.
Thus, we can consider that the fourth-order perturbation theory well describes the partial breakdown of the FL state.
In order to more accurately describe the breakdown of the FL state, we require higher-order terms.
We can easily imagine that such higher-order terms will become important in the shaded area in the low-frequency region in Fig.~\ref{fig:Mott}.
The difference from the simple atomic limit is the remaining dispersive behavior in the low-frequency region.

\begin{figure}[t!]
\begin{center}
\includegraphics[height=65mm]{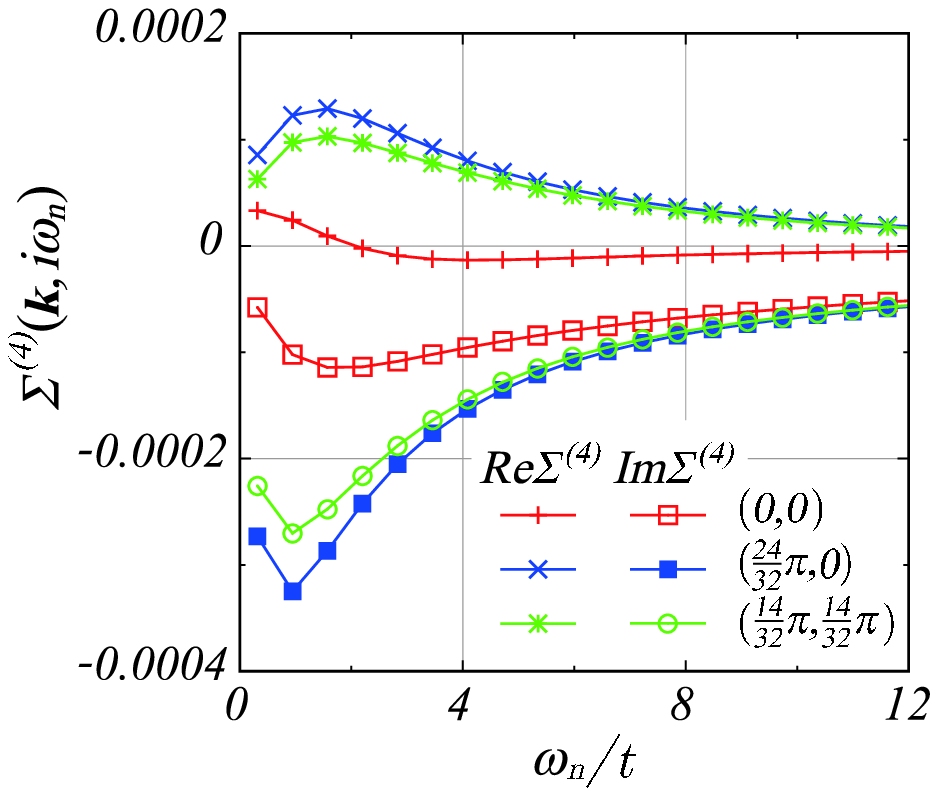}
\caption{(Color online) Fourth-order terms $\mathit{\Sigma}^{(4)}({\mib k},\mathrm{i}\omega_n)$ at ${\mib k}=(0,0)$, $(\frac{24}{32}\pi,0)$ and $(\frac{14}{32}\pi,\frac{14}{32}\pi)$ for $(T,\mu)=(0.1t,-0.72t)$ in the doped case.
The magnitude of the imaginary part is about 3 times smaller than the maximum value at ${\mib k}=(\pi,0)$ in the half-filled case.
It decreases in the low-frequency region, which is the conventional FL behavior.}
\label{fig:sig4thn090}

\vspace{10 pt}
\includegraphics[height=65mm]{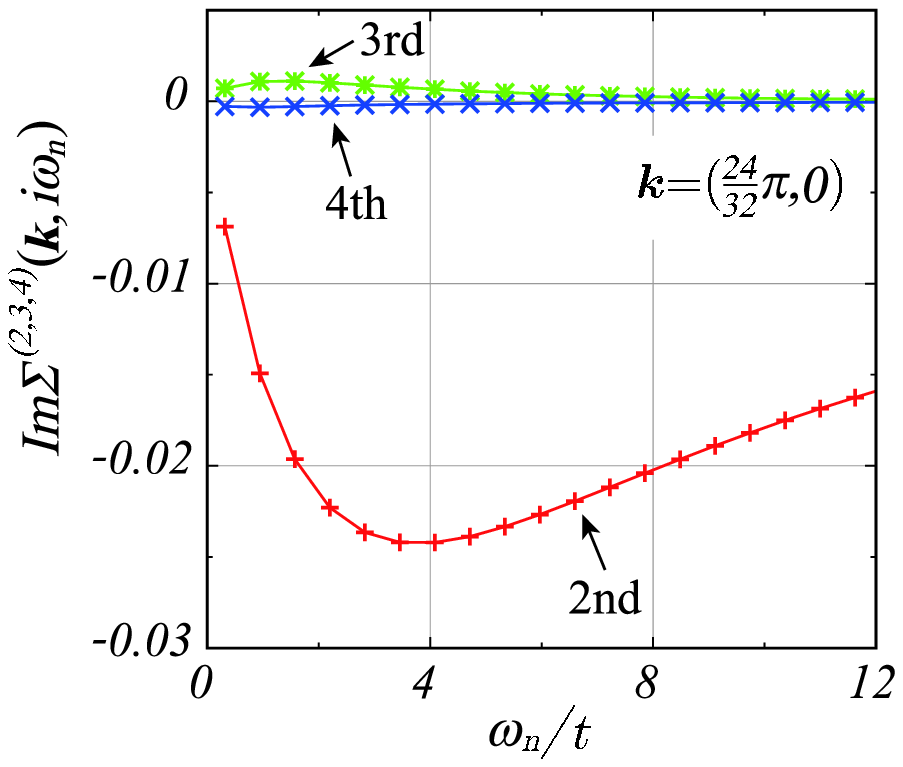}
\caption{(Color online) Imaginary part of the $n$th order terms for $n=2$, $3$, $4$ at ${\mib k}=(\frac{24}{32}\pi,0)$ in the doped case.
One can see that the third- and fourth-order terms are much smaller than the second-order term.}
\label{fig:sign090}
\end{center}
\end{figure}
{\it Doped case} \rule[2pt]{24pt}{0.5pt}
Next, let us consider a case except that for the half-filling.
For instance, $\mu=-0.72t$ corresponds to $n\simeq 0.9$ at $U=6t$.
The system possesses neither particle-hole symmetry nor the complete nesting.
In this case, the FL state is considered more stable than that in the half-filled case.
We can expect that the fourth-order perturbation is valid in a wider parameter region.
In fact, we indicate that the perturbation expansion has better convergence, and thus, is applicable even at lower temperatures than that in the half-filled case.

\begin{figure}[t!]
\begin{center}
\includegraphics[height=65mm]{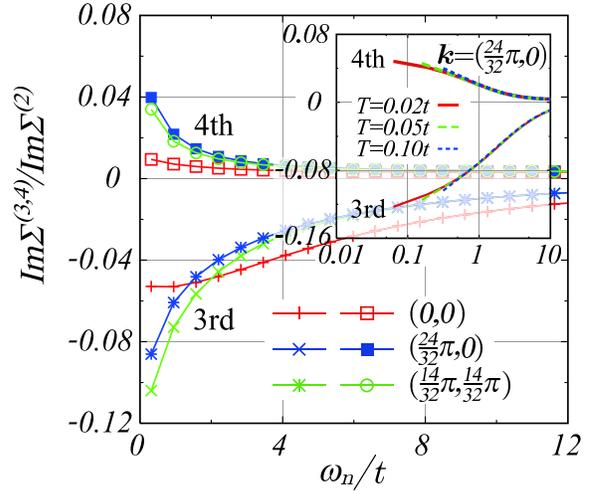}
\caption{(Color online) Ratios of $\mathrm{Im}\mathit{\Sigma}^{(3)}({\mib k},\mathrm{i}\omega_n)$ and $\mathrm{Im}\mathit{\Sigma}^{(4)}({\mib k},\mathrm{i}\omega_n)$ to $\mathrm{Im}\mathit{\Sigma}^{(2)}({\mib k},i\omega_n)$ at ${\mib k}=(0,0)$, $(\frac{24}{32}\pi,0)$ and $(\frac{14}{32}\pi,\frac{14}{32}\pi)$ in the doped case.
The ratio of the third order is negative.
This is because only the third-order term possesses positive values, as shown in Fig.~\ref{fig:sign090}.
The ratios of the third and fourth orders are, respectively, smaller than $\frac{1}{3!}/\frac{1}{2!}=1/3$ and $\frac{1}{4!}/\frac{1}{2!}=1/12$.
The convergence of the perturbation expansion is considered very good.
The inset denotes the ratios at $(\frac{24}{32}\pi,0)$ for $T=0.02t$, $0.05t$ and $0.1t$ as a function of $\log(\omega_n/t)$.
The maximum values do not become very large even for $T\lesssim 0.1t$.}
\label{fig:ration090}
\end{center}
\end{figure}

In Fig.~\ref{fig:sig4thn090}, we display the $\omega_n$ dependence of the fourth-order term at two points near the FS, ${\mib k}=(\frac{24}{32}\pi,0)$ and $(\frac{14}{32}\pi,\frac{14}{32}\pi)$, and the $\mathit{\Gamma}ma$ point.
The magnitude of the imaginary part $\mathrm{Im}\mathit{\Sigma}^{(4)}({\mib k},\mathrm{i}\omega_n)$ is larger than that of the real part $\mathrm{Re}\mathit{\Sigma}^{(4)}({\mib k},\mathrm{i}\omega_n)$, and has the largest value at ${\mib k}=(\frac{24}{32}\pi,0)$.
The value is about 3 times smaller than the largest value at ${\mib k}=(\pi,0)$ in the half-filled case.
In addition, it turns to decrease in the low-frequency region.
This is the conventional FL behavior, different from the behavior at ${\mib k}=(\pi,0)$ in the half-filled case.
In Fig.~\ref{fig:sign090}, we illustrate the imaginary part of each order term at ${\mib k}=(\frac{24}{32}\pi,0)$.
In this case, where the system is asymmetric with respect to the electron-hole, the third-order term $\mathit{\Sigma}^{(3)}(k)$ does not vanish and has a finite value.
One can see that the third- and fourth-order terms are much smaller than the second-order term.
This indicates that the perturbation expansion converges rapidly.
In Fig.~\ref{fig:ration090}, we evaluate the ratios of $\mathrm{Im}\mathit{\Sigma}^{(3)}({\mib k},\mathrm{i}\omega_n)$ and $\mathrm{Im}\mathit{\Sigma}^{(4)}({\mib k},\mathrm{i}\omega_n)$ to $\mathrm{Im}\mathit{\Sigma}^{(2)}({\mib k},\mathrm{i}\omega_n)$ quantitatively.
The difference in sign between these ratios comes from the fact that only the third-order term possesses a positive sign, as shown in Fig.~\ref{fig:sign090}.
This implies that the third-order term works to make the effective mass lighter.
Thus, the third-order perturbation theory is insufficient for the effective mass.
We need the fourth-order perturbation to obtain a large mass enhancement factor.
In Fig.~\ref{fig:ration090}, we can see that the ratio of the fourth order is suppressed abruptly in the high-frequency region, and becomes almost zero like that in the half-filled case.
However, strictly speaking, the ratio approaches a small but finite value, different from that in the half-filled case.
This is because, in the doped case, the second-order term does not provide the exact form at the atomic limit, and generally, we require higher-order terms.

Now, let us discuss the convergence of the perturbation series.
If we expect the convergence of $\sim 1/n!$ as discussed in the half-filled case, the ratios of the third- and fourth-order terms have to be of the order of $\frac{1}{3!}/\frac{1}{2!}=1/3$ and $\frac{1}{4!}/\frac{1}{2!}=1/12$, respectively.
The actual ratios are smaller than those values.
The convergence of the perturbation expansion is considered very good.
In particular, the maximum ratio of the fourth-order term is smaller than that in the half-filled case, and does not become very large even at lower temperatures, as shown in the inset of Fig.~\ref{fig:ration090}.
This indicates that the perturbation expansion is valid even for $T\lesssim 0.1t$, and the FL state is more stable than that in the half-filled case.
This is because the doped system possesses neither the strong AF fluctuation by the complete nesting nor the Mott-Hubbard transition.

However, we require some cautions in this consideration.
First, we can take $\rho_0 U$ as the expansion parameter, not the present $U/t$,
where $\rho_0$ is the DOS at the Fermi level.
In this case, the ratio in the half-filled case becomes relatively smaller than that in the doped case, since $\rho_0$ is largest in the half-filled case.
Then the convergence of the perturbation expansion may almost be the same regardless of the electron filling.
Second, we cannot simply apply the convergence criterion to the perturbation expansion in the present calculation.
This is because the third-order term is relatively small in the vicinity of the half-filling since it completely vanishes just at half-filling.
Therefore, we examined the ratio of the fourth-order term to the second-order term, not the ratio of consecutive order terms.
We need to evaluate higher-order terms to discuss the convergence of the perturbation series more accurately, although this is very difficult.

{\it Validity of the fourth-order perturbation} \rule[2pt]{24pt}{0.5pt}
From the above results, we would like to stress that the coefficient of the fourth-order term in the perturbation expansion in $U$ is very small as compared with that of the second-order term.
For $T\gtrsim 0.1t$ in the half-filled case, the maximum ratio is smaller than $\sim 1/12(=\frac{1}{4!}/\frac{1}{2!})$, and in the doped case, it remains smaller even for $T\lesssim 0.1t$.
Such behavior of the coefficients is very similar to that of the specific heat coefficient in the impurity Anderson model.~\cite{rf:Zlatic2}
Also in that case, the odd-number order terms vanish in the symmetric case, and the coefficient of each $n$th-order term rapidly decreases roughly in proportion to $1/n!$.
The smallness of the fourth-order coefficient in the lattice system partially proves such good convergence.
Thus, we can expect that the perturbation expansion in $U$ maintains good convergence in a wide parameter space of $(n,T)$ also in the lattice system, except for $T \lesssim 0.1t$ in the half-filled case.
In order to further clarify the behavior of $\sim 1/n!$, we need to investigate the coefficient of the sixth-order term, at least.
It is one of important future works, but difficult to perform.
However, we consider that such good convergence in the perturbation expansion is inevitable, following the concept of the adiabatic continuation mentioned in \S 1.~\cite{rf:Anderson}
As long as no phase transition occurs, the system connects adiabatically with the noninteracting system, and the physical quantities are analytic with respect to $U$.

Now, let us consider the validity of the perturbation expansion truncated within a finite order.
We examine the range of $U$ where the second-order perturbation theory is valid, by comparing the second-order term and the fourth-order term, since the third-order term vanishes in the half-filled case.
At $T=0.1t$ in the half-filled case, the second-order perturbation is quantitatively valid for $U \lesssim 3t\sim 4t$ from the ratio in Fig.~\ref{fig:ratio}.
In the doped case, it holds even for $T\lesssim 0.1t$ from the inset in Fig.~\ref{fig:ration090}.
For $T \gtrsim 0.1t$, the second-order perturbation is also valid for larger $U$.
In the same way, to examine the validity of the fourth-order perturbation, we need the sixth-order term.
In the present situation, we cannot discuss the validity quantitatively, since we do not estimate the contribution of the sixth-order term.
However, if the convergence of the perturbation expansion is good as we expected, and the coefficient of each $n$th-order term rapidly decreases roughly in proportion to $1/n!$, like that in the impurity case, then the fourth-order perturbation can be considered quantitatively valid for $U \lesssim 5t \sim 6t$ from $u^4/4! \simeq u^6/6!$.
From the smallness of the fourth-order coefficient and the concept of the adiabatic continuation mentioned above, we can expect that the expansion coefficients behave like $\sim 1/n!$ also in the lattice system, although we cannot guarantee it since we cannot estimate the correct contribution of neglected higher-order terms.
In the following section, we display the DOS in the fourth-order perturbation for $U<10t$ at $T=0.1t$ in the half-filled case.
It shows a striking feature for $U>5t$.
We believe that the asymptotic behavior for large $U$ in the fourth-order perturbation is well worth studying, even though it cannot be validated at present.
Probably, the feature obtained in the fourth-order perturbation will become more distict by the inclusion of higher-order terms.

\begin{figure}[t!]
\begin{center}
\vspace{10 pt}
\includegraphics[height=6cm]{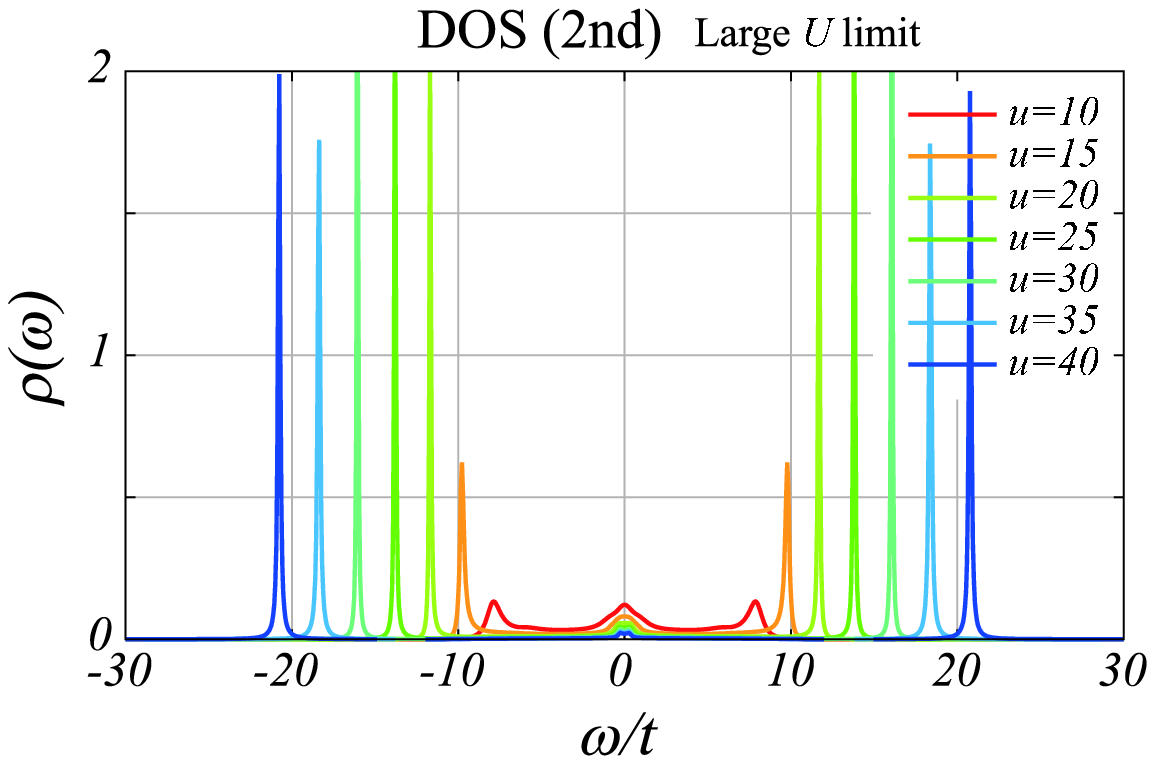} \\
\caption{(Color online) DOS in the second-order perturbation at large $u=U/t$ limit at $T=0.1t$ in the half-filled case.
It exhibits two peaks like the {$\delta$ function} at around $\omega\simeq\pm U/2$ for $U\gtrsim 30t$.
This is consistent with the fact that the second-order perturbation is exact at the atomic limit $t\to 0$.
In this case, the DOS is given by $\rho(\omega)=\frac{1}{2}\bigl(\delta(\omega+\frac{U}{2})+\delta(\omega-\frac{U}{2})\bigr)$.}
\label{fig:dosinc}

\vspace{20 pt}
\includegraphics[height=6cm]{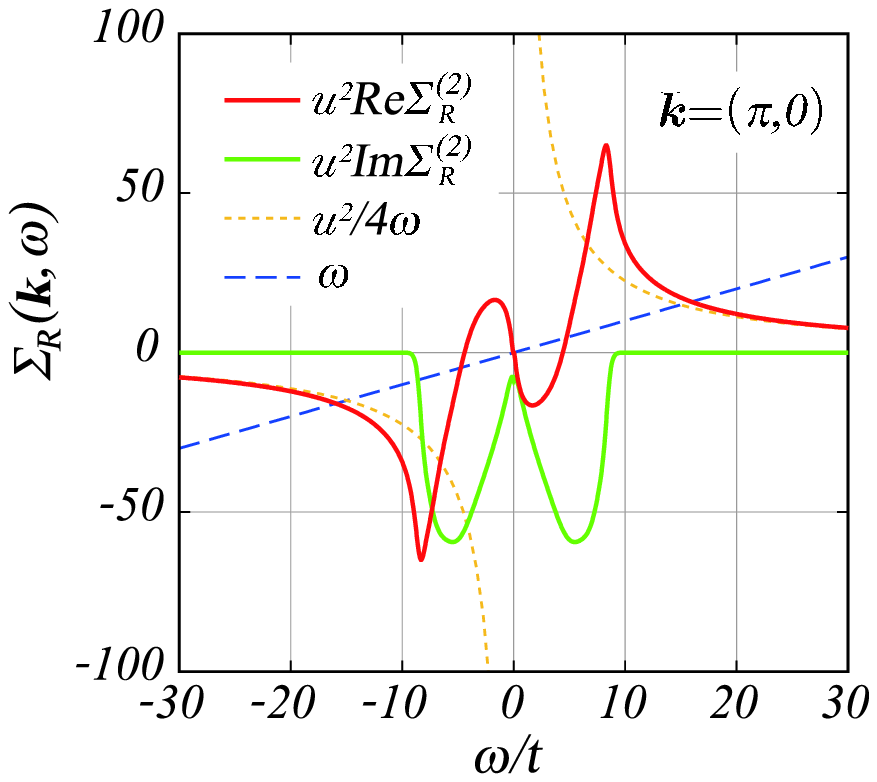}
\caption{(Color online) Second-order retarded self-energy at ${\mib k}=(\pi,0)$ for $U=30t$ in the half-filled case.
It asymptotically approaches the atomic limit $U^2/4\omega$ in the high energy region.
The real part of the self-energy crosses $\omega$ at around $\omega \simeq \pm U/2=\pm 15t$, at which the imaginary part is negligible.
Thus, the retarded Green's function possesses the real poles at the intersections, and the DOS exhibits the {$\delta$ function} peaks there.}
\label{fig:siginc}
\end{center}
\end{figure}
\subsection{Density of States}

We go on to a study of the DOS.
In this section, we confine ourselves to the half-filled case.
First, we review the DOS in the second-order perturbation and the FLEX.
Then, we explain the result of the fourth-order perturbation.
The DOS in each approximation is evaluated by carrying out ${\mib k}$ summation of Green's function with the corresponding self-energy, and then using analytic continuation.
That is, the DOS is given by
\begin{equation}
\rho(\omega)=-\frac{1}{\pi}\mathrm{Im}\mathcal{G}_R({\mib r}=0,\omega),
\end{equation}
where $\mathcal{G}_R({\mib r}=0,\omega)$ is the analytic continuation of
\begin{equation}
\mathcal{G}({\mib r}=0,\mathrm{i}\omega_n)=\frac{1}{N^2}\sum_{\mib k}\mathcal{G}({\mib k},\mathrm{i}\omega_n).
\end{equation}
The analytic continuation on the real axis is numerically calculated 
with the use of the Pad\'e approximation.~\cite{rf:Vidberg}

\begin{figure}[t!]
\begin{center}
\vspace{10 pt}
\includegraphics[height=6cm]{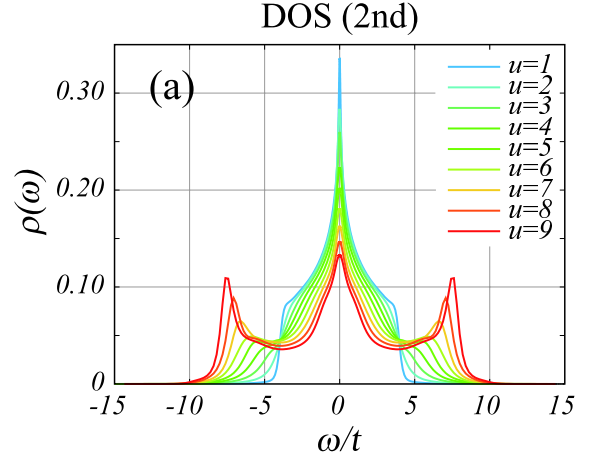} \\
\vspace{10 pt}
\includegraphics[height=6cm]{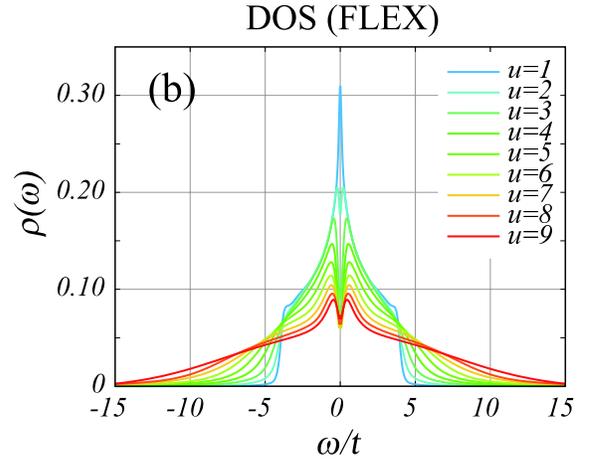} \\
\vspace{10 pt}
\includegraphics[height=6cm]{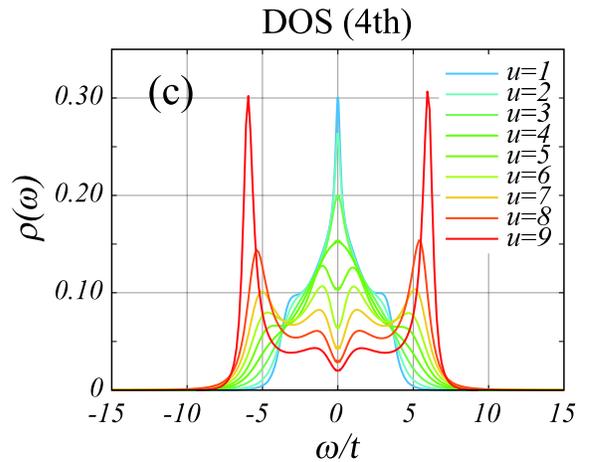}
\caption{(Color online) (a) DOSs in the second-order perturbation, (b) FLEX and (c) fourth-order perturbation at $T=0.1t$ in the half-filled case.
In (a), we can see the three-peak structure, which corresponds to the upper and lower Hubbard bands, and the coherent quasiparticle central peak.
In (b), the pseudogap behavior at the Fermi level appears owing to the strong AF fluctuation, although the incoherent Hubbard structure is smeared and unclear.
In (c), we find distinct incoherent peaks and the pseudogap structure for large $U\gtrsim 5t$.
This structure is considered as the precursor of the Mott-Hubbard AF structure.}
\label{fig:dos}
\end{center}
\end{figure}
{\it Second-order perturbation} \rule[2pt]{24pt}{0.5pt}
In Fig.~\ref{fig:dos}(a), we exhibit the DOS in the second-order perturbation.
The characteristic three-peak structure corresponds to the upper and lower Hubbard bands, and the coherent quasiparticle central peak.
With increasing $U$, the structure becomes distinct.
The width of the central peak corresponds to the quasiparticle bandwidth.
With larger $U$, it becomes narrower, and the spectral weight is transferred to the incoherent Hubbard peaks in the high energy region.
As shown in Fig.~\ref{fig:dosinc}, at the large $U$ limit, the quasiparticle weight almost vanishes, and the Hubbard peaks become similar to the {$\delta$ function}.
The DOS substantially behaves like $\rho(\omega)\simeq \frac{1}{2}\bigl(\delta(\omega+\frac{U}{2})+\delta(\omega-\frac{U}{2})\bigr)$.
This can be understood from the behavior of the retarded self-energy $\mathit{\Sigma}_R({\mib k},\omega)=u^2\mathit{\Sigma}^{(2)}_R({\mib k},\omega)$ for large $U=30t$ in Fig.~\ref{fig:siginc}.
In this case, the real part crosses $\omega$ at around $\omega\simeq\pm U/2=\pm 15t$, at which the imaginary part is negligible.
In other words, the retarded Green's function $\mathcal{G}_R({\mib k},\omega)$ possesses real poles at around $\omega\simeq\pm U/2$.
Such behavior in the second-order perturbation is reasonable, not an artifact of the approximation, since it provides the exact form of self-energy at the large $U$ limit, that is, the atomic limit ($t\to 0$).~\cite{rf:Hewson}
This is a special circumstance in the half-filled case.
In this case, in the atomic limit, the exact retarded self-energy is given by $U^2/4\omega$, and the Green's function possesses two poles at $\omega=\pm U/2$.
In Fig.~\ref{fig:dosinc}, we can see that the DOS almost restores such behavior for $U\gtrsim 30t$, although the peak positions are still slightly larger.

{\it FLEX} \rule[2pt]{24pt}{0.5pt}
In Fig.~\ref{fig:dos}(b), we display the DOS in the FLEX for various $U$ values.
The pseudogap occurs at the Fermi level $\omega=0$ with a larger $U$ owing to the strong AF fluctuation.
On the other hand, the incoherent Hubbard bands in the high energy region are smeared and unclear.
This is because the FLEX does not properly include the effect of the local correlation, as mentioned in \S 1.
If we are reminded of our discussion of Fig.~\ref{fig:Mott}, this is related to the fact that the FLEX is inconsistent with the second-order perturbation in the high-frequency region, that is, it does not restore the atomic limit.

{\it Fourth-order perturbation} \rule[2pt]{24pt}{0.5pt}
Finally, the DOS in the fourth-order perturbation is illustrated in Fig.~\ref{fig:dos}(c).
For $U<3t$, the behavior is almost the same as that in the second-order perturbation.
It shows a qualitatively similar behavior even at $U=4t$.
For $U>5t$, however, the shape is quite different from that in the second-order perturbation and the FLEX.
It exhibits distinct incoherent peaks and the growth of the pseudogap at the Fermi level.
The full-width of the DOS shrinks as compared with the second-order perturbation.
This is considered as the band-narrowing effect caused by the correlation.
In this case, the incoherent Hubbard bands are located at around the atomic limit positions already at $U\simeq W$, where $W=8t$ is the bandwidth.
This value is much smaller than $U\simeq 30t$ in the second-order perturbation.
This is because the self-energy in the fourth-order perturbation is closer to the atomic limit form, as shown in Fig.~\ref{fig:Mott}.
The pseudogap develops with larger $U$, and the energy scale, $\sim 2t$, is much larger than that in the FLEX.
Such a structure in the DOS is rather similar to that in the DCA,~\cite{rf:Maier} although the peak structure of the incoherent part is sharper in the fourth-order perturbation.
For $U=W$, the DCA has shown the formation of two coherent bands above and below $\omega=0$ owing to the strong AF correlation in addition to broad upper and lower Hubbard bands.
This characteristic Mott-Hubbard AF four band structure has been discussed by the quantum Monte Carlo method~\cite{rf:Maier,rf:Preuss,rf:Bulut} and the extension of DMFT~\cite{rf:Kusunose}.
Here, we have found that the DOS in the fourth-order perturbation also exhibits similar behavior.

Thus, we can consider that the fourth-order perturbation appropriately describes the asymptotic behavior into the Mott-Hubbard transition.
In order to clarify such behavior, let us next investigate the behavior of self-energy at ${\mib k}$ points on the FS.

\begin{figure}[t!]
\begin{center}
\vspace{10 pt}
\includegraphics[height=6cm]{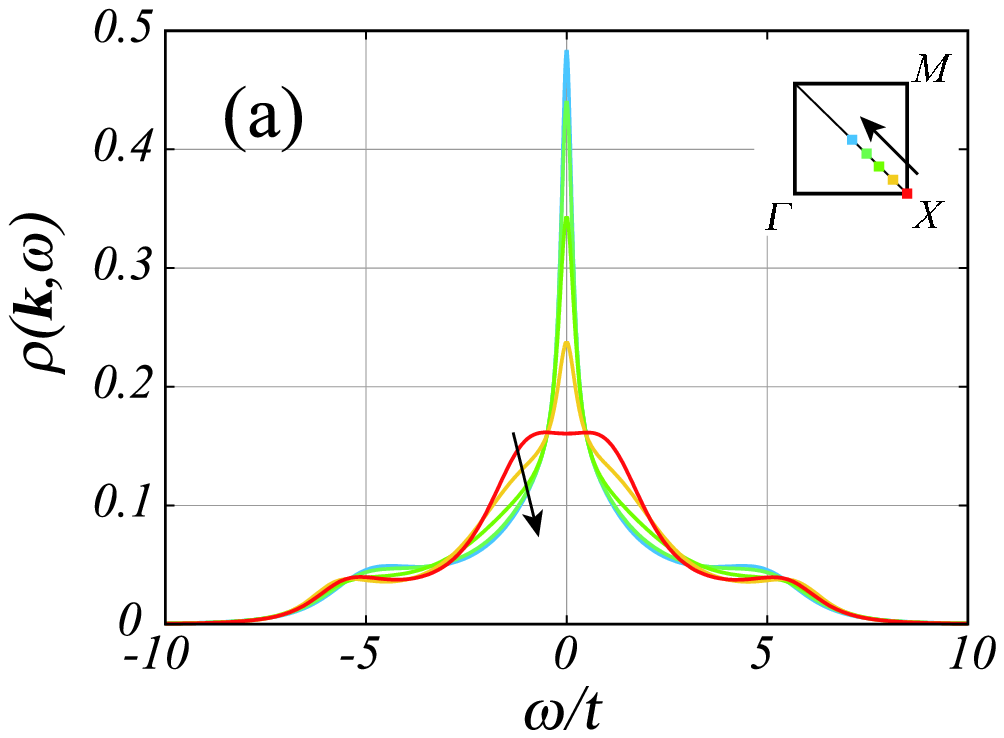} \\
\vspace{10 pt}
\includegraphics[height=6cm]{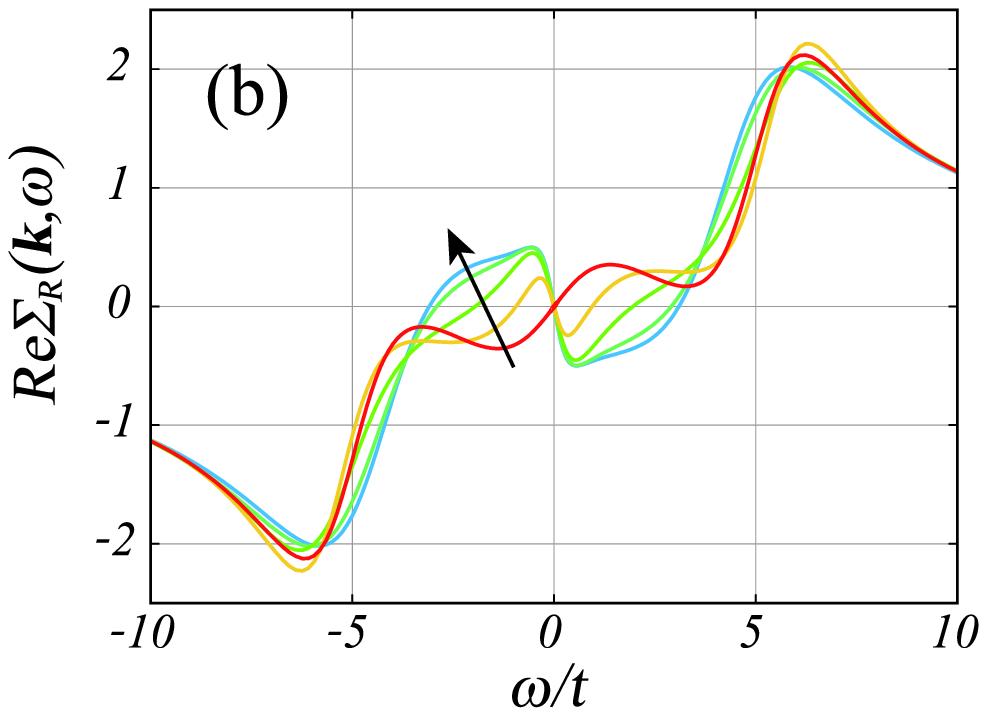} \\
\vspace{10 pt}
\includegraphics[height=6cm]{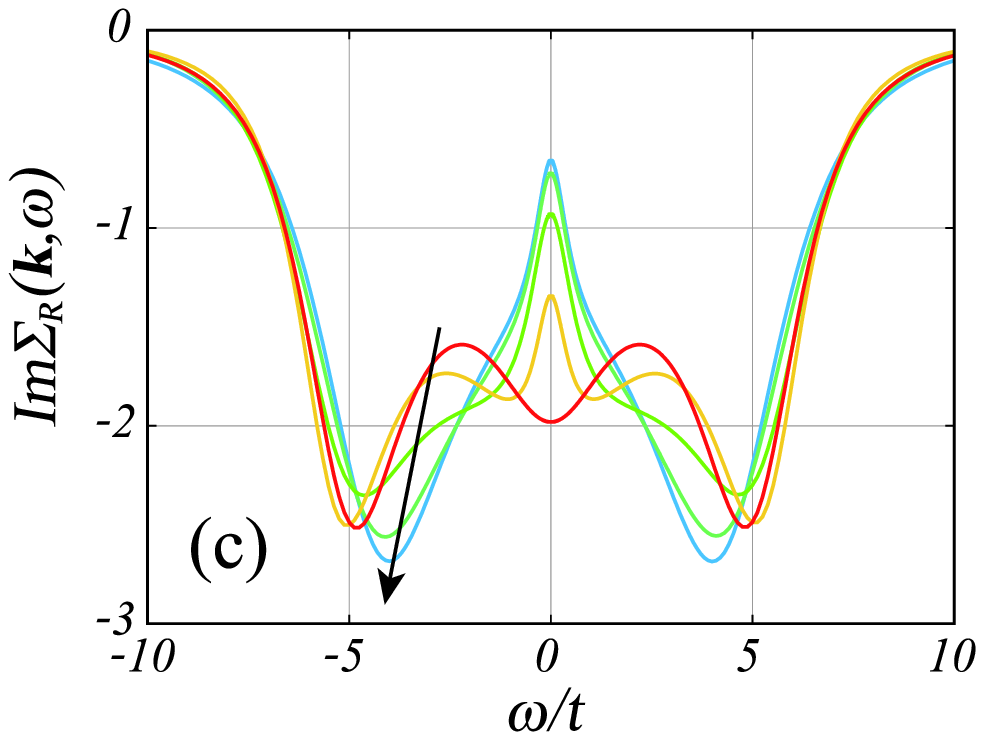}
\caption{(Color online) (a) Spectral weight, (b) the real part of the retarded self-energy and (c) the imaginary part for $U=6t$ at several ${\mib k}$ points on the FS shown in the inset of (a) in the half-filled case.
The quasiparticle peak is well defined at ${\mib k}=(\pi/2,\pi/2)$, far from which it gradually decreases, and then disappears at ${\mib k}=(\pi,0)$.
Correspondingly, the real part of the retarded self-energy at ${\mib k}=(\pi,0)$ has a positive slope at $\omega=0$, at which the imaginary part indicates a large negative.
Such behavior is quite different from the conventional FL behavior.}
\label{fig:rhoU06}
\end{center}
\end{figure}
\subsection{Spectral weight and self-energy}

Figure \ref{fig:rhoU06} represents the spectral weight
\begin{equation}
\rho({\mib k},\omega)=-\frac{1}{\pi}\mathrm{Im}\mathcal{G}_R({\mib k},\omega)
\end{equation}
and the retarded self-energy $\mathit{\Sigma}_R({\mib k},\omega)$ at several ${\mib k}$ points on the FS for $U=6t$ in the half-filled case.
Note that, as shown in Fig.~\ref{fig:rhoU06}(a), the spectral weight at ${\mib k}=(\pi,0)$ does not exhibit the coherent quasiparticle peak structure, although that at ${\mib k}=(\pi/2,\pi/2)$ maintains the characteristic three peak structure.
This indicates that the quasiparticle is not well defined at ${\mib k}=(\pi,0)$, although it still survives at ${\mib k}=(\pi/2,\pi/2)$.
In fact, the real part of $\mathit{\Sigma}_R({\mib k},\omega)$ at ${\mib k}=(\pi,0)$ in Fig.~\ref{fig:rhoU06}(b) has a positive slope at $\omega=0$, at which the imaginary part in Fig.~\ref{fig:rhoU06}(c) has a large negative value.
Such behavior is quite different from the conventional FL behavior observed at ${\mib k}=(\pi/2,\pi/2)$.
Thus, the strong correlation breaks down the conventional FL quasiparticle at ${\mib k}=(\pi,0)$.
It may be rather better to consider that new two coherent bands are formed owing to the AF correlation, since the real part of the retarded self-energy has a negative slope at around $\omega\simeq\pm 2t$ and the magnitude of the imaginary part is relatively small.
In fact, the spectral weight seems to exhibit new hump structures at around $\omega \simeq \pm t$.~\cite{rf:Spec}
This is consistent with the formation of the Mott-Hubbard AF four-band structure mentioned above.~\cite{rf:Maier,rf:Kusunose,rf:Preuss,rf:Bulut}
In this case, however, far from ${\mib k}=(\pi,0)$, the structure is gradually suppressed, and the quasiparticle peak is restored.
The pseudogap is large at ${\mib k}=(\pi,0)$, and vanishes at ${\mib k}=(\pi/2,\pi/2)$.
This is different from the fully gapped behavior in the Mott-Hubbard AF structure.
Although the DOS should be fully gapped at large $U$ limit, in the fourth-order perturbation, even for larger $U$, the pseudogap is not open at ${\mib k}=(\pi/2,\pi/2)$.
As discussed in Fig.~\ref{fig:Mott}, such gap behavior is related to the tendency toward an increase in the imaginary part of the self-energy $\mathrm{Im}\mathit{\Sigma}({\mib k},\mathrm{i}\omega_n)$ with a small $\omega_n$.
$\mathrm{Im}\mathit{\Sigma}({\mib k},\mathrm{i}\omega_n)$ at ${\mib k}=(\pi/2,\pi/2)$ does not show the tendency even for large $U=13t$.
The situation will be improved with the higher-order terms, since they are expected to possess a large contribution in the shaded area in Fig.~\ref{fig:Mott}.
Thus, the pseudogap behavior in the fourth-order perturbation can be regarded as the precursor of the Mott-Hubbard AF four band structure.

\begin{figure}[t!]
\begin{center}
\vspace{10 pt}
\includegraphics[height=65mm]{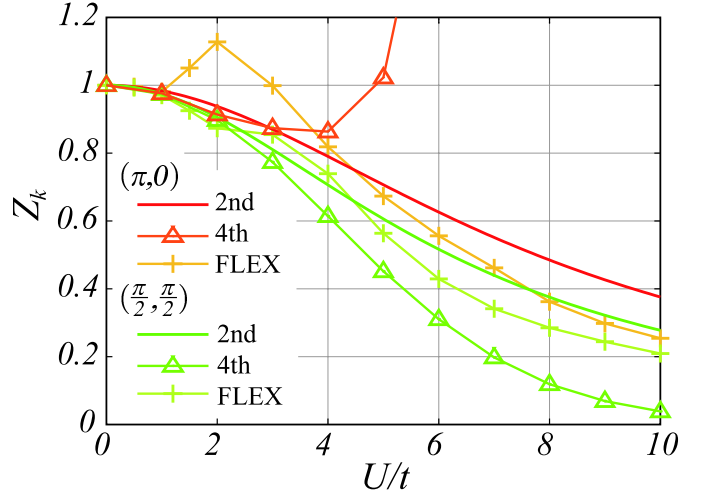}
\caption{(Color online) Mass renormalization factor $z_{\mib k}$ at ${\mib k}=(\pi,0)$ and $(\pi/2,\pi/2)$ in the half-filled case.
In the second-order perturbation, $z_{\mib k}$ is given by $1/(1+u^2 0.0166)$ at ${\mib k}=(\pi,0)$ and $1/(1+u^2 0.026)$ at ${\mib k}=(\pi/2,\pi/2)$.
In the FLEX, $z_{\mib k}$ is more suppressed than that in the second order.
In the fourth-order perturbation, $z_{\mib k}$ at ${\mib k}=(\pi/2,\pi/2)$ has the smallest value.
This means the formation of a quasiparticle with very heavy mass.
On the other hand, $z_{\mib k}$ at ${\mib k}=(\pi,0)$ becomes negative at $U>6t$.
This corresponds to the formation of the Mott-Hubbard AF structure.}
\label{fig:zk}
\end{center}
\end{figure}
\subsection{Mass enhancement factor}

The mass enhancement factor, which is the inverse of the mass renormalization factor $z_{\mib k}$, is given by the slope of the real part of the retarded self-energy at $\omega=0$,
\begin{equation}
z_{\mib k}^{-1}=1-\frac{\partial \mathrm{Re}\mathit{\Sigma}_R({\mib k},\omega)}{\partial \omega}\bigg|_{\omega \to 0}.
\end{equation}
In Fig.~\ref{fig:zk}, we illustrate the $U$ dependence of $z_{\mib k}$ at ${\mib k}=(\pi,0)$ and $(\pi/2,\pi/2)$ in the half-filled case.
In the second-order perturbation, we obtain $z_{\mib k}^{-1}=1+u^2 0.0166$ at ${\mib k}=(\pi,0)$ and $z_{\mib k}^{-1}=1+u^2 0.026$ at ${\mib k}=(\pi/2,\pi/2)$ from the analytic continuation of $\mathit{\Sigma}^{(2)}({\mib k},\mathrm{i}\omega_n)$.
In the FLEX and the fourth-order perturbation, we carry out the Pad\'e approximation for self-energy at each $U$.
Concerning the quasiparticle at ${\mib k}=(\pi/2,\pi/2)$, which is always well defined, $z_{\mib k}$ in the fourth-order perturbation is the smallest among the three approximations, and is very small for a large $U$.
This indicates the formation of the quasiparticle with very heavy mass.
The mass enhancement factor $z_{\mib k}^{-1}$ in the fourth-order perturbation is several times larger than that in the FLEX.
We can expect that it becomes larger at lower temperatures.
Such behavior is also obtained for the generic case except in the case of half-filling.
Thus, the fourth-order perturbation theory can describe the quasiparticle with mass as heavy as that in heavy fermion systems.
On the other hand, at ${\mib k}=(\pi,0)$, the fourth-order perturbation is qualitatively different from the other two approximations.
In the fourth-order perturbation, for $U>6t$, $z_{\mib k}$ possesses a negative value and the quasiparticle is not well defined.
This corresponds to the formation of the Mott-Hubbard AF structure as mentioned above.
Although we cannot deny that this may mean the breakdown of the perturbation expansion, 
it will be clarified by studying higher-order terms.

\begin{figure}[t!]
\begin{center}
\vspace{10 pt}
\includegraphics[height=6cm]{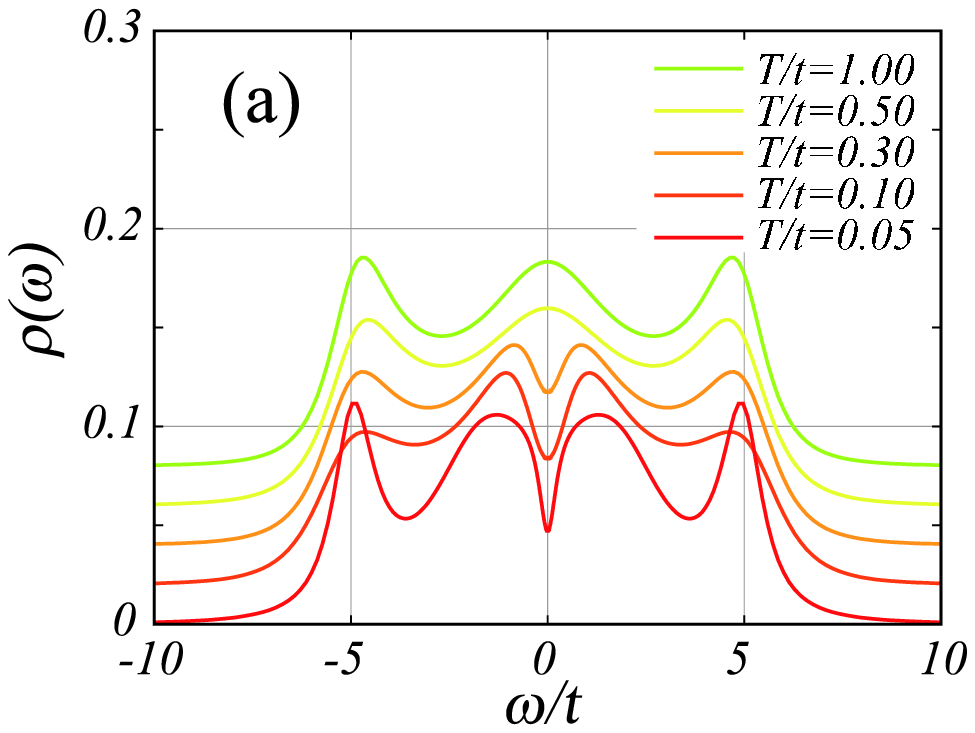} \\
\vspace{10 pt}
\includegraphics[height=6cm]{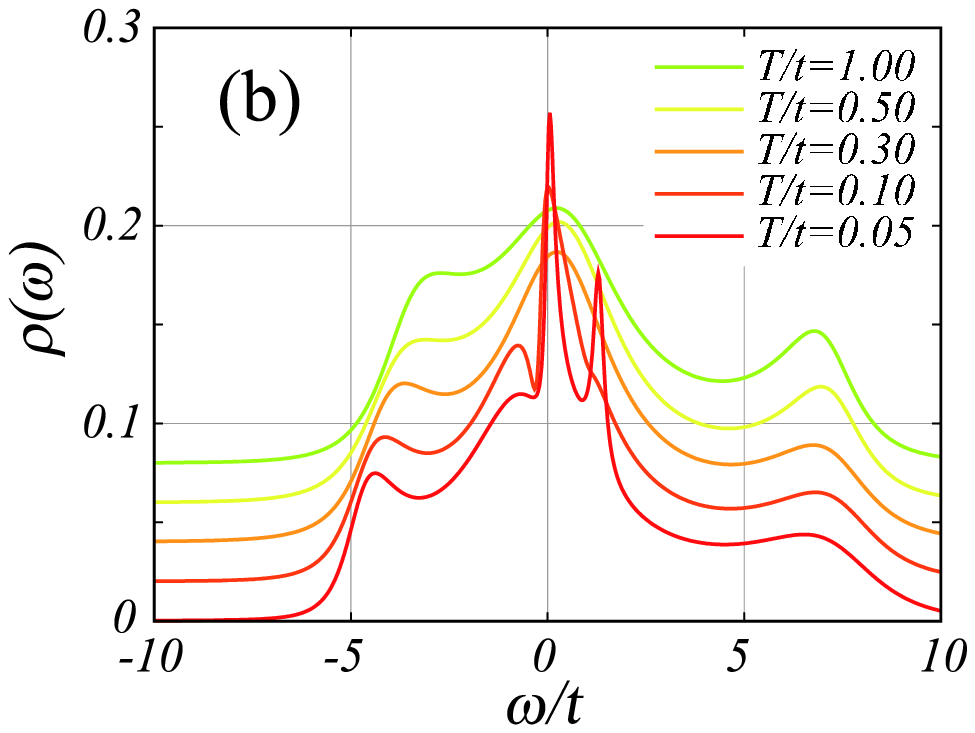}
\caption{(Color online) DOSs at $U=6t$ (a) in the half-filled case $n=1$ and (b) doped case $n \simeq 0.9$.
The origin of the vertical axis is shifted by $0.02$ for each temperature.
As shown in (a), with decreasing temperature, the pseudogap at the Fermi level develops.
This indicates the asymptotic behavior when the FL state breaks down.
In contrast, in (b), we can see that the sharp peak at the Fermi level grows.
Thus, in this case, a FL quasiparticle with heavy mass is formed.}
\label{fig:dosT}
\end{center}
\end{figure}
\subsection{Temperature dependence}

Finally, let us discuss the temperature dependence of the DOS in the fourth-order perturbation.
In Fig.~\ref{fig:dosT}, we illustrate the DOS for various temperatures at $U=6t$.
Figures \ref{fig:dosT}(a) and \ref{fig:dosT}(b) show results in the half-filled case $n=1$ and doped case $n\simeq 0.9$, respectively.
In Fig.~\ref{fig:dosT}(b), the quasiparticle peak at the Fermi level and the side peaks develop with decreasing temperature.
The behavior of the former is consistent with the conventional FL state, although the latter are too enhanced due to the Pad\'e approximation.
Thus, in the doped case, the FL quasiparticles with heavy mass are formed with decreasing temperature.
On the other hand, in Fig.~\ref{fig:dosT}(a), the pseudogap develops at the Fermi level with decreasing temperatures.
This corresponds to the formation of the Mott-Hubbard AF structure as discussed above.
We cannot find such behavior for small $U$ in the same temperature region.
Thus, the fourth-order perturbation expansion seems to describe the asymptotic behavior into the Mott-Hubbard transition.
This unexpected result is very fascinating.
This characteristic behavior will also be reflected in transport phenomena.
It is very interesting to clarify whether the conductivity remains metallic.
If it shows insulating behavior or a precursor, then we can study how to approach the metal-insulator transition.
In order to clarify such behavior, we need further investigation.
All these will be investigated in our next study.

\section{Summary and Discussion}
In this study, we have investigated the Hubbard model on a two-dimensional square lattice by the perturbation expansion to the fourth order in the on-site Coulomb repulsion $U$.
Numerically calculating all diagrams up to the fourth order in self-energy, we have examined the convergence of the perturbation series in the lattice system.
The numerical techniques used here will be useful in calculating generic vertex correction terms, which cannot be reduced to their convolution form.
In addition, we have reported the results of the DOS and the mass enhancement factor.
Let us here summarize our results below.

First, one of the most important results is the fact that the coefficient of the fourth-order term in self-energy is much smaller than that of the second-order term.
The ratio is large at a small $\omega_n$, and the maximum values at $T=0.1t$ are $\sim 0.07$ in the half-filled case and $\sim 0.04$ in the doped case.
These values are consistent with the behavior of $\sim 1/n!$ for the $n$th-order term, which is almost the same as the behavior of the expansion coefficients in the impurity Anderson model.~\cite{rf:Zlatic2}
Such good convergence in the perturbation expansion is considered as a general feature in the FL state from the concept of the adiabatic continuation.~\cite{rf:Anderson}
The smallness of the fourth-order coefficient partially proves such good convergence in the lattice system, although we cannot guarantee it since we cannot exactly evaluate neglected higher-order terms.
Thus, we can expect that, as far as the FL state is stable, the lattice system also maintains good convergence, and the coefficients of higher-order terms are very small.
In fact, the fourth-order term at a large $\omega_n$ almost vanishes.
This is consistent with the fact that the FL state is fairly stable at high temperatures.
In addition, for $T\lesssim 0.1t$, although the maximum ratio does not become very large in the doped case, it becomes large rapidly in the half-filled case.
This is probably because the ground state at $T=0$ is the Mott-Hubbard AF state in the half-filled case, although the FL state is stable in the doped case.
Thus, in the present lattice system, we can expect that the perturbation expansion in $U$ keeps good convergence in wide parameter space of $(n,T)$, except for $T\lesssim 0.1t$ in the half-filled case.

Next, we evaluated the range of validities of the perturbation expansion truncated within a finite order.
We can examine the range of $U$ where the second-order perturbation theory is valid, by comparing the second-order term and the fourth-order term, since the third-order term vanishes in the half-filled case.
From the above ratios, the second-order perturbation proves to be quantitatively valid for $U\lesssim 3t\sim 4t$.
We cannot exactly determine the range of validity of the fourth-order perturbation, since we did not estimate the sixth-order term.
However, following the $1/n!$-like behavior mentioned above, we can expect that it is quantitatively valid for $U\lesssim 5t\sim 6t$ from $u^4/4!\simeq u^6/6!$.

As for the behavior in the high-frequency region, the fourth-order term is abruptly suppressed owing to almost perfect cancellation of contributions from each diagram.
At half-filling, in particular, the second-order term provides the exact form of the self-energy in the high-frequency region.
Thus, the self-energy to the fourth order keeps the atomic limit form $U^2/4\mathrm{i}\omega_n$ in the high-frequency region.~\cite{rf:Hewson}
This is important for the incoherent Hubbard peaks and the feature of the Mott transition.
The failure in the FLEX comes partially from the fact that it is inconsistent with the atomic limit form in the high-frequency region.
In the fourth-order perturbation, the self-energy for a large $U$ more closely approaches the atomic limit form.
This indicates that it can reflect the Mott-Hubbard characters more strongly, which have not been grasped in the second-order perturbation so far.
In fact, we found the asymptotic breakdown of the FL state in the strong correlation regime as the pseudogap behavior in the DOS.
In order to clarify this feature more precisely, we need to evaluate higher-order terms.
We can expect that those terms decrease more rapidly in the high-frequency region, and most of the contributions are confined in the low-frequency region.

Furthermore, we found interesting behaviors in the DOS.
At half-filling, two features are noticeable.
One is about the incoherent Hubbard structure in the high energy region.
The peak structure is more remarkable, and the position is already very close to the atomic limit value $\omega\simeq\pm U/2$ for $U\simeq W$, where $W$ is the bandwidth $8t$, although the second-order perturbation barely restores such behavior at around $U\simeq 4W$.
Another is about the remarkable pseudogap phenomena at the Fermi level.
With increasing $U$ or decreasing temperature, the pseudogap develops.
The energy scale $\sim 2t$ is much larger than that of the simple AF gap observed in the FLEX.
The pseudogap is rather considered as the precursor of the formation of the Mott-Hubbard AF structure~\cite{rf:Maier,rf:Kusunose,rf:Preuss,rf:Bulut}
In this case, although the spectral weight at around ${\mib k}=(\pi,0)$ opens the pseudogap, the structure is gradually suppressed far from ${\mib k}=(\pi,0)$, and the quasiparticle peak is restored at around ${\mib k}=(\pi/2,\pi/2)$.
This is different from the conventional isotropic AF gap.
Although the DOS should be fully gapped at a large $U$ limit, the pseudogap remains close at ${\mib k}=(\pi/2,\pi/2)$ in the fourth-order perturbation.
This implies that higher-order terms are required at a larger $U$.
Rather, such gapless behavior in the fourth-order perturbation seems to be consistent with the Fermi arc phenomena observed in the ARPES.~\cite{rf:Shen}
However, this should be discussed with a more realistic band structure and electron filling. This is one of our future works.
On the other hand, in the doped case, the DOS exhibits both a narrow quasiparticle peak at the Fermi level and an incoherent Hubbard structure.
In this case, we obtain a very large mass enhancement factor.
Thus, the fourth-order perturbation theory overall well explains the asymptotic behavior in the strong correlation regime.

Finally, let us suggest several future works.
The pseudogap behavior obtained in the fourth-order perturbation begins to develop in a rather high-temperature region.
In this case, it is very interesting to clarify whether the resistivity remains metallic.
The study of transport phenomena based on the perturbation expansion is one of our future works.
Another work is about AF and superconducting transitions within the fourth-order perturbation.
We can evaluate the superconducting (or AF) gap equation in the fourth-order perturbation theory.
The preliminary results for the superconducting transition indicate that with increasing $U$, the eigenvalue of d-wave spin-singlet pairing increases abruptly in the weak correlation regime, and becomes optimal at around $U\simeq 5t$, and then gradually decreases in the strong correlation regime.
Such behavior is reasonable, and is also consistent with the result of recent variational Monte Carlo calculations.~\cite{rf:Yokoyama,rf:Yokoyama2}
The increase of eigenvalue with increasing $U$ originates from the increase of the attractive force, and the decrease of eigenvalue for a large $U$ limit results mainly from the increase of the mass enhancement factor, namely, the renormalization effect.~\cite{rf:Shinkai,rf:Shinkai2}
Thus, the fourth-order perturbation theory throws light on investigations of the moderate correlation regime, which have been very difficult to perform so far.
The phase diagram and physical properties will be discussed in our forthcoming paper.

\section*{Acknowledgements}
One of the authors (H.I.) thanks S. Fujimoto for valuable discussions.
The numerical calculations were carried out on SX8 at YITP in Kyoto University.
This study is financially supported by a Grant-in-Aid for Scientific Research 
on Priority Areas (Grant No. 18043016) from the Ministry of Education, 
Culture, Sports, Science and Technology of Japan.

\section*{Appendix}
In the symmetric case, the third-order term vanishes, and three terms contained in each type of the fourth-order term becomes equivalent.
These exact relations can also be a test to check numerical calculations.
Let us here prove these relations.

The dispersion relation given in eq.~(\ref{eq:Disp}) satisfies
$\epsilon_{{\mib k}+{\mib Q}}=-\epsilon_{\mib k}$ with ${\mib Q}=(\pi,\pi)$.
At half-filling $n=1$, the chemical potential $\mu=0$.
In this case, the single-particle Green's function possesses the property
\begin{align}
\mathcal{G}_0(k+Q)
&=\mathcal{G}_0({\mib k}+{\mib Q},\mathrm{i}\omega_n) \nonumber \\[-1mm]
&=\frac{1}{\mathrm{i}\omega_n+\epsilon_{\mib k}} \nonumber \\
&=-\mathcal{G}_0(-k), \nonumber
\end{align}
where $Q=({\mib Q},0)$.
The ladder diagram is related to the bubble diagram,
\begin{align}
\phi(q+Q)
&=\sum_k \mathcal{G}_0(k)\mathcal{G}_0(q-k+Q) \nonumber \\[-1mm]
&=-\sum_k \mathcal{G}_0(k)\mathcal{G}_0(k-q) \nonumber \\
&=\chi(q). \nonumber
\end{align}
Thus, 
\begin{align}
\sum_q\phi_0(q)^2\mathcal{G}_0(q-k)
&=\sum_q\phi_0(q+Q)^2\mathcal{G}_0(q-k+Q) \nonumber \\[-1mm]
&=-\sum_q\chi_0(q)\mathcal{G}_0(k-q), \nonumber
\end{align}
and then, the third-order term is proved to vanish.

Next, we consider each term of the fourth order.
First of all, since the second-order self-energy is calculated as
\begin{align}
\mathit{\Sigma}^{(2)}(k+Q)
&=-\sum_q\phi_0(q)\mathcal{G}_0(q-k-Q) \nonumber \\[-1mm]
&=-\sum_q\phi_0(q+Q)\mathcal{G}_0(q-k) \nonumber \\[-1mm]
&=-\sum_q\chi_0(q)\mathcal{G}_0(q-k) \nonumber \\
&=-\mathit{\Sigma}^{(2)}(-k), \nonumber
\end{align}
then we obtain
\[
\mathcal{G}_1(k+Q)=-\mathcal{G}_1(-k).
\]
Thus, in the self-energy correction term $\mathit{\Sigma}^{(4)}_\mathrm{self}(k)$,
\begin{align}
-\sum_q\phi_0(q)\mathcal{G}_1(q-k)
&=-\sum_q\phi_0(q+Q)\mathcal{G}_1(q+Q-k) \nonumber \\[-1mm]
&= \sum_q\chi_0(q)\mathcal{G}_1(k-q), \nonumber
\end{align}
and then, $(a)-(c)$ in Fig.~\ref{fig:sig4} have equivalent contributions.
In the RPA term $\mathit{\Sigma}^{(4)}_\mathrm{RPA}(k)$,
\begin{align}
-\sum_q\phi_0^3(q)\mathcal{G}_0(q-k)
&=-\sum_q\phi_0^3(q+Q)\mathcal{G}_0(q+Q-k) \nonumber \\[-1mm]
&= \sum_q\chi_0^3(q)\mathcal{G}_0(k-q), \nonumber 
\end{align}
and then, $(d)-(f)$ in Fig.~\ref{fig:sig4} have equivalent contributions.
Next, from $\Lambda''(k+Q,q)=\Lambda(k,q)$,
\begin{align}
\chi_1'(q)
&=\Lambda''(p+Q,q)\mathcal{G}_0(-p-Q)\mathcal{G}_0(q-p-Q) \nonumber \\
&=\Lambda(p,q)\mathcal{G}_0(p)\mathcal{G}_0(p-q) \nonumber \\
&=\chi_1(q), \nonumber
\end{align}
and from $\Lambda'(k,q+Q)=-\Lambda(k,q)$,
\begin{align}
\phi_1(q+Q)
&=\Lambda'(p,q+Q)\mathcal{G}_0(p)\mathcal{G}_0(q-p+Q) \nonumber \\
&=\Lambda(p,q)\mathcal{G}_0(p)\mathcal{G}_0(p-q) \nonumber \\
&=\chi_1(q). \nonumber
\end{align}
Thus, in the type-I vertex correction term $\mathit{\Sigma}^{(4)}_\mathrm{vtx1}(k)$,
\begin{align}
-\sum_q\chi_1(q)\mathcal{G}_0(k-q)
&=-\sum_q\chi_1'(q)\mathcal{G}_0(k-q) \nonumber \\[-1mm]
&= \sum_q\phi_1(q)\mathcal{G}_0(q-k), \nonumber
\end{align}
that is, $(g)-(i)$ in Fig.~\ref{fig:sig4} have equivalent contributions.
Finally, in the type-II vertex correction term $\mathit{\Sigma}^{(4)}_\mathrm{vtx2}(k)$,
\begin{align}
\begin{split}
&\sum_q\Lambda'(k,q)\phi_0(q)\mathcal{G}_0(q-k) \\[-1mm]
&\hspace{10mm} =\sum_q\Lambda'(k,q+Q)\phi_0(q+Q)\mathcal{G}_0(q-k+Q)
\end{split}
\nonumber \\[-1mm]
&\hspace{10mm} =\sum_q\Lambda(k,q)\chi_0(q)\mathcal{G}_0(k-q). \nonumber
\end{align}
Namely, $(j)-(l)$ in Fig.~\ref{fig:sig4} have equivalent contributions.



\begin{thebibliography}{99}
\bibitem{rf:Spec}
The difference between the characteristic energy of the spectral weight ($\simeq \pm t$) and that of the retarded self-energy ($\simeq \pm 2t$) results from the renormalization effect.
\end{thebibliography}
\end{document}